\documentclass[lettersize,journal]{IEEEtran}

\usepackage{amsmath,amsfonts,amssymb}
\usepackage{algorithmic}
\usepackage{algorithm}
\usepackage{array}
\usepackage[caption=false,font=normalsize,labelfont=sf,textfont=sf]{subfig}
\usepackage{textcomp}
\usepackage{stfloats}
\usepackage{url}
\usepackage{verbatim}
\usepackage{graphicx}
\usepackage{cite}
\usepackage{cite}
\usepackage{url}								
\usepackage{siunitx}							
\usepackage{booktabs}							
\usepackage{xltabular}
\usepackage{tabularx}
\newcolumntype{Y}{>{\centering\arraybackslash}X} 
\newcolumntype{Z}{>{\raggedleft\arraybackslash}X} 
\usepackage{enumitem} 
\usepackage{makecell}							
\usepackage{multirow}							
\usepackage[display-foreign=false]{acro}			
\usepackage{gensymb}
\usepackage{upgreek}							
\acsetup{single}								
\usepackage{mathtools}
\usepackage{bm}
\usepackage{yhmath}
\usepackage{physics}
\usepackage{xfrac}
\usepackage{jabbrv}						

\usepackage{filecontents}
\usepackage{xr-hyper}
\usepackage{color}
\usepackage{xcolor}

\DeclareAcronym{si}{
	short = SI,
	long = International System of Units,
}
\DeclareAcronym{csi}{
	short = CSI,
	long = channel state information,
}
\DeclareAcronym{rfe}{
	short = RFFE,
	long = radio frequency front-end,
}
\DeclareAcronym{cwtt}{
	short = CWTT,
	long = continuous-wave two-tone,
}
\DeclareAcronym{ipc}{
	short = IPC,
	long = interprocess communication,
}
\DeclareAcronym{awg}{
	short = AWG,
	long = arbitrary waveform generator,
}
\DeclareAcronym{tbp}{
	short = TBP,
	long = time--bandwidth product,
}
\DeclareAcronym{cfo}{
	short = CFO,
	long = carrier frequency offset,
}\DeclareAcronym{sfo}{
	short = SFO,
	long = sampling frequency offset,
}

\DeclareAcronym{fpga}{
	short = FPGA,
	long = field-programmable gate array,
}
\DeclareAcronym{gpu}{
	short = GPU,
	long = graphics processing unit,
}

\DeclareAcronym{nlos}{
	short = NLoS,
	long = non-line-of-sight,
}
\DeclareAcronym{ptt}{
	short = PTT,
	long = pulsed two-tone,
}
\DeclareAcronym{tdma}{
	short = TDMA,
	long = time division multiple access,
}
\DeclareAcronym{tof}{
	short = ToF,
	long = time of flight,
	long-plural = times of flight,
}
\DeclareAcronym{toa}{
	short = ToA,
	long = time of arrival,
	long-plural = times of arrival,
}
\DeclareAcronym{imd}{
	short = IMD,
	long = intermodulation distortion,
}
\DeclareAcronym{fm}{
	short = FM,
	long = frequency modulation,
}
\DeclareAcronym{pm}{
	short = PM,
	long = phase modulation,
}

\DeclareAcronym{xo}{
	short = XO,
	long = crystal oscillator,
}
\DeclareAcronym{tcxo}{
	short = TCXO,
	long = temperature compensated \acl{xo},
}
\DeclareAcronym{ocxo}{
	short = OCXO,
	long = oven controlled \acl{xo},
}

\DeclareAcronym{fir}{
	short = FIR,
	long = finite impulse response
}
\DeclareAcronym{dsp}{
	short = DSP,
	long = digital signal processor
}
\DeclareAcronym{ls}{
	short = LS,
	long = least-squares
}
\DeclareAcronym{qls}{
	short = QLS,
	long = quadratic least-squares
}
\DeclareAcronym{sinc-ls}{
	short = sinc-LS,
	long = sinc nonlinear least-squares
}
\DeclareAcronym{mf-ls}{
	short = MFLS,
	long = matched filter least-squares
}
\DeclareAcronym{rf}{
	short = RF,
	long = radio frequency
}
\DeclareAcronym{lfm}{
	short = LFM,
	long = linear frequency modulation
}
\DeclareAcronym{prf}{
	short = PRF,
	long = pulse repetition frequency
}
\DeclareAcronym{pri}{
	short = PRI,
	long = pulse repetition interval
}
\DeclareAcronym{fmcw}{
	short = FMCW,
	long = frequency modulated continuous-wave
}
\DeclareAcronym{lfmcw}{
	short = LFMCW,
	long = linear frequency modulated continuous-wave
}
\DeclareAcronym{cw}{
	short = CW,
	long = continuous-wave
}
\DeclareAcronym{dbf}{
	short = DBF,
	long = digital beamforming
}
\DeclareAcronym{sar}{
	short = SAR,
	long = synthetic aperture radar
}
\DeclareAcronym{psr}{
	short = PSR,
	long = point scatterer response
}
\DeclareAcronym{rcs}{
	short = RCS,
	long = radar cross-section
}
\DeclareAcronym{crlb}{
	short = CRLB,
	long = Cram\'er-Rao lower bound
}
\DeclareAcronym{dof}{
	short = DoF,
	long = degree of freedom
}
\DeclareAcronym{snr}{
	short = SNR,
	long = signal-to-noise ratio
}
\DeclareAcronym{sinr}{
	short = SINR,
	long = signal-to-interference-plus-noise ratio
}
\DeclareAcronym{fft}{
	short = FFT,
	long = fast Fourier transform,
}
\DeclareAcronym{ift}{
	short = IFT,
	long = inverse Fourier transform,
}
\DeclareAcronym{rmse}{
	short = RMSE,
	long = root-mean-square error
}
\DeclareAcronym{psd}{
	short = PSD,
	long = power spectral density
}
\DeclareAcronym{rca}{
	short = RCA,
	long = range of closest approach
}
\DeclareAcronym{rda}{
	short = RDA,
	long = Range-Doppler Algorithm
}
\DeclareAcronym{rma}{
	short = RMA,
	long = Range Migration Algorithm
}
\DeclareAcronym{pfa}{
	short = PFA,
	long = Polar Formatting Algorithm
}
\DeclareAcronym{bpa}{
	short = BPA,
	long = Backprojection Algorithm
}
\DeclareAcronym{rvp}{
	short = RVP,
	long = residual video phase
}
\DeclareAcronym{jrc}{
	short = JRC,
	long = joint radar-communications
}
\DeclareAcronym{doa}{
	short = DOA,
	long = direction of arrival
}
\DeclareAcronym{hci}{
	short = HCI,
	long = human-computer interaction
}
\DeclareAcronym{its}{
	short = ITS,
	long = intelligent transportation systems
}
\DeclareAcronym{rtk}{
	short = RTK,
	long = real-time kinematic
}
\DeclareAcronym{eirp}{
	short = EIRP,
	long = effective isotropic radiated power
}
\DeclareAcronym{gnss}{
	short = GNSS,
	long = global navigation satellite system
}
\DeclareAcronym{imu}{
	short = IMU,
	long = inertial measurement unit
}

\DeclareAcronym{ofdm}{
	short = OFDM,
	long = orthogonal frequency division multiplexing
}
\DeclareAcronym{los}{
	short = LoS,
	long = line of sight
}

\DeclareAcronym{pll}{
	short = PLL,
	long = phase-locked loop
}
\DeclareAcronym{vco}{
	short = VCO,
	long = voltage-controlled oscillator
}
\DeclareAcronym{lna}{
	short = LNA,
	long = low-noise amplifier
}
\DeclareAcronym{if}{
	short = IF,
	long = intermediate frequency,
	short-indefinite = an,
	long-indefinite = an
}
\DeclareAcronym{cots}{
	short = COTS,
	long = commercial off-the-shelf
}
\DeclareAcronym{adc}{
	short = ADC,
	long = analog to digital converter
}
\DeclareAcronym{dac}{
	short = DAC,
	long = digital to analog converter
}
\DeclareAcronym{lo}{
	short = LO,
	long = local oscillator
}
\DeclareAcronym{pcb}{
	short = PCB,
	long = printed circuit board
}
\DeclareAcronym{mimo}{
	short = MIMO,
	long = multiple-input multiple-output
}
\DeclareAcronym{simo}{
	short = SIMO,
	long = single-input multiple-output
}
\DeclareAcronym{mmic}{
	short = MMIC,
	long = monolithic microwave integrated circuit
}
\DeclareAcronym{daq}{
	short = DAQ,
	long = data acquisition
}
\DeclareAcronym{ic}{
	short = IC,
	long = integrated circuit
}
\DeclareAcronym{pa}{
	short = PA,
	long = power amplifier
}

\DeclareAcronym{ti}{
	short = TI,
	long = Texas Instruments
}
\DeclareAcronym{adi}{
	short = ADI,
	long = Analog Devices
}

\DeclareAcronym{roi}{
	short = ROI,
	long = region of interest,
	long-plural-form = regions of interest
}
\DeclareAcronym{v2x}{
	short = V2X,
	long = vehicle-to-everything
}
\DeclareAcronym{av}{
	short = AV,
	long = automated vehicle
}
\DeclareAcronym{cors}{
	short = CORS,
	long = continuously operating reference station
}
\DeclareAcronym{mdot}{
	short = MDOT,
	long = Michigan Department of Transportation
}
\DeclareAcronym{moco}{
	short = MOCO,
	long = motion compensation
}
\DeclareAcronym{sdr}{
	short = SDR,
	long = software-defined radio
}
\DeclareAcronym{gpio}{
	short = GPIO,
	long = general-purpose input/output
}
\DeclareAcronym{usrp}{
	short = USRP,
	long = Universal Software Radio Peripheral
}
\DeclareAcronym{uhd}{
	short = UHD,
	long = \ac{usrp} Hardware Driver
}
\DeclareAcronym{ntp}{
	short = NTP,
	long = network time protocol
}
\DeclareAcronym{ptp}{
	short = PTP,
	long = precision time protocol
}
\DeclareAcronym{lan}{
	short = LAN,
	long = local area network
}
\DeclareAcronym{wlan}{
	short = WLAN,
	long = wireless \ac{lan}
}
\DeclareAcronym{lut}{
	short = LUT,
	long = lookup table
}
\DeclareAcronym{wsn}{
	short = WSN,
	long = wireless sensor network
}
\DeclareAcronym{mac}{
	short = MAC,
	long = media access control
}
\DeclareAcronym{pps}{
	short = PPS,
	long = pulse-per-second
}
\DeclareAcronym{fom}{
	short = FoM,
	long = figure of merit
}
\DeclareAcronym{uwb}{
	short = UWB,
	long = ultra-wideband
}
\DeclareAcronym{twtt}{
	short = TWTT,
	long = two-way time transfer
}
\DeclareAcronym{adas}{
	short = ADAS,
	long = advanced driver assistance systems
}
\DeclareAcronym{swap}{
	short = SWaP,
	long = {size, weight, and power}
}
\DeclareAcronym{cda}{
	short = CDA,
	long = {coherent distributed antenna array}
}
\DeclareAcronym{ap}{
	short = AP,
	long = {access point}
}

\DeclareMathOperator{\sinc}{sinc}

\def\Vrulefillleft#1#2{
    \leavevmode%
    \hskip-2ex%
    \leaders%
    \vtop{\hsize=.001in\vskip#1#2}%
    \hfill%
    \hskip2ex%
}
\def\Vrulefillright#1#2{
    \leavevmode%
    \hskip1ex%
    \leaders%
    \vtop{\hsize=.001in\vskip#1#2}%
    \hfill%
    \hskip-0.5ex%
}

\newcommand*\centeredline[1]{{\color{lightgray}\Vrulefillleft{-0.8ex}{\_}}#1{\color{lightgray}\Vrulefillright{-0.8ex}{\_}}}

\makeatletter
\def\@firstoftwo@second#1#2{%
  \def\temp##1.##2\@nil{##2}%
   \temp#1\@nil}
\newcommand\sref[1]{%
   \expandafter\@setref\csname r@#1\endcsname\@firstoftwo@second{#1}%
}

\newcommand{\hlb}[1]{#1}

\begin{document}


\title{Real-Time High-Accuracy Digital Wireless Time, Frequency, and Phase Calibration for Coherent Distributed Antenna Arrays}

\author{Jason M. Merlo,~\IEEEmembership{Member,~IEEE,} Samuel Wagner, John B. Lancaster, and Jeffrey A. Nanzer,~\IEEEmembership{Senior Member,~IEEE}
\thanks{Manuscript received June 00, 2025. This work was supported in part under the auspices of the U.S. Department of Energy by Lawrence Livermore National Laboratory under Contract DEAC52-07NA27344, by the LLNL LDRD Program under Project No. 22-ER-035 and 25-ER-040, in part by the Office of Naval Research under award \#N00014-25-1-2208, and in part by the National Science Foundation under grant \#1751655. Distribution Statement A. Approved for public release: distribution unlimited. Release number: LLNL-JRNL-2006754.}
\thanks{J. M. Merlo and J. A. Nanzer are with the Department of Electrical and Computer Engineering, Michigan State University, East Lansing, MI 48824 USA (e-mail: merlojas@msu.edu; nanzer@msu.edu).}
\thanks{S. Wagner and J. B. Lancaster are with the Lawrence Livermore National Laboratory, Livermore, CA 94550 USA.}
\thanks{Supplemental materials for this paper are available on IEEE Xplore.}
}


\maketitle

\begin{abstract}
This work presents a fully-digital high-accuracy real-time calibration procedure for frequency and time alignment of open-loop wirelessly coordinated \ac{cda} modems, enabling \ac{rf} phase coherence of spatially separated \ac{cots} \acp{sdr} without cables or external references such as \ac{gnss}. Building on previous work using high-accuracy spectrally-sparse \ac{toa} waveforms and a multi-step \ac{toa} refinement process, a high-accuracy \ac{twtt}-based time--frequency coordination approach is demonstrated. 
Due to the two-way nature of the high-accuracy \ac{twtt} approach, the time and frequency estimates are Doppler and multi-path tolerant, so long as the channel is reciprocal over the synchronization epoch. This technique is experimentally verified using \ac{cots} \acp{sdr} in a lab environment in static and dynamic scenarios and with significant multipath scatterers. Time, frequency, and phase stability were evaluated by beamforming over coaxial cables to an oscilloscope which achieved time and phase precisions of $\sim$$\mathbf{\SI{60}{\pico\second}}$--$\mathbf{\SI{70}{\pico\second}}$, with median coherent gains above 99\,\% using optimized coordination parameters, and a beamforming frequency \ac{rmse} of 3.73\,ppb in a dynamic scenario. Finally, experiments were conducted to compare the performance of this technique with previous works using an analog \ac{cwtt} frequency reference technique in both static and dynamic settings.
\end{abstract}
\acresetall


\begin{figure}
	\centering
	\includegraphics[width=0.9\columnwidth]{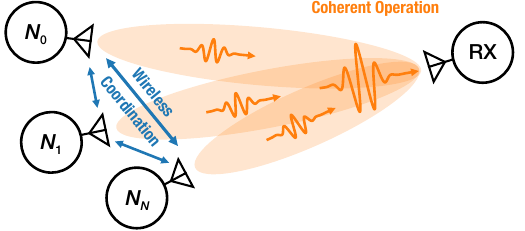}	
	\caption{Distributed array concept for an $N$-element array. Each element in the array coordinates wirelessly to align its electrical state (time, frequency, and phase) to perform a coherent operation, such as beamforming. In this example, the receiver, labelled RX, can be a conventional active receiver modem, or a passive feature of interest in a radar-type operation.}
	\label{fig:ov1}
\end{figure}

\section{Introduction}

\IEEEPARstart{W}{irelessly coordinated} \acp{cda}, e.g., Fig. \ref{fig:ov1}, have been growing in interest in a range of applications spanning from radar imaging and remote sensing~\cite{prager2020wireless,schlegel2023coherent-distri,merlo2024ims,aguilar2024uncoupled-digital-radars,merlo2024aps,kenney2024distributed-frequency-and-phase-synchronization,kenney2025concept-and-the,nusrat2024distributed-repeaters,werbunat2024synchronization-of-radars,shen2024high-resolution}, automotive radar~\cite{tagliaferri2021cooperative-syn,janoudi2024signal-model-fo,werbunat2024multichannel-re,fenske2025integrated-self}, and wireless power transfer~\cite{choi2018distributed-wir,brunet2024wireless-power-transfer,mahbub2024far-field-wirel}, to secure wireless communications~\cite{ellison2022distributed}, next generation terrestrial wireless communications~\cite{holtom2024discobeam,bondada2024wireless-mobile,xu2024enhancement-of-,sun2022secure-and-ener}, and deep space communication networks~\cite{quadrelli2019distributed-swarm},~\cite[TX05.2.6]{nasa2020taxonomy}. The motivation for this growing interest in wireless \acp{cda}, instead of a traditional monolithic array, is driven by several factors: scalability, adaptability, and reliability~\cite{nanzer2021distributed-arrays}. Wirelessly coordinated \acp{cda} can be easily scaled to satisfy mission requirements, and nodes can be easily replaced if they should fail.  Because the nodes are mobile, the array geometry can adapt to meet changing requirements. Finally, because the functions of the \ac{cda} can be spread across many nodes, there is no single point of failure should an element become inoperable.

To achieve these benefits, the electrical states---time, phase, and frequency---of each system must be aligned to ensure all elements operate coherently. This can be accomplished by either providing feedback from a cooperative destination, forming a \textit{closed-loop} system, e.g., \cite{brown2005method,seo2008feedback,brown-iii2008time-slotted-ro,mudumbai2009distributed-tra,preuss2011two-way-synchro,bidgare2012implementation,quitin2013a-scalable-arch,hanna2023distributed-transmit-beamforming-uav}, or by communicating the electrical states directly between nodes in the array, forming an \textit{open-loop} system~\cite{nanzer2017open-loop-coher}, e.g., \cite{abari2015airshare,yan2019software-define,ouassal2020decentralized,alemdar2021rfclock,ouassal2021decentralized,prager2020wireless,mghabghab2021open-loop-distr,merlo2023tmtt,hanna2023distributed-transmit-beamforming,werbunat2024synchronization-of-radars,kenney2024distributed-frequency-and-phase-synchronization,kenney2024frequency-synchronization,holtom2024discobeam,shandi2024decentralized-p,merlo2024past,ghasemi2024time-frequency-synchronization}. While closed-loop topologies are often simpler to implement, they suffer from the fact that individual elements must still be able to receive feedback from the destination, requiring sufficient single-element \ac{snr}; furthermore, closed-loop systems inherently require feedback from the receiver, and thus, cannot support communication with conventional modems or radar remote sensing operations. Open-loop \acp{cda} can support any operating mode that a conventional antenna array can support, but are more challenging to implement as the exact location, time, frequency, and phase offsets of all nodes must be coordinated directly between elements to a fraction of the carrier wavelength to support fully-coherent operation~\cite{nanzer2017open-loop-coher}. To accomplish this, a standard timebase for the array which describes the current time and frequency must be agreed upon. In general, there are two methods of selecting a timebase in an array: \textit{centralized} in which single node is elected to act as the primary timebase that disseminates its time and frequency to all other nodes in the array, e.g.,~\cite{abari2015airshare,yan2019software-define,mghabghab2021open-loop-distr,alemdar2021rfclock,merlo2023tmtt,hanna2023distributed-transmit-beamforming,werbunat2024synchronization-of-radars,holtom2024discobeam}, or \textit{decentralized} in which all elements in the array collectively converge to a mean time and frequency value, e.g.,~\cite{ouassal2020decentralized,ouassal2021decentralized,kenney2024distributed-frequency-and-phase-synchronization,kenney2024frequency-synchronization}; however, in some cases, a combination of these approaches is employed~\cite{prager2020wireless,shandi2024decentralized-p}. Typically, in centralized arrays, a star topology is used, where all secondary nodes receive information directly from the primary node---repeaters can be employed but will degrade the coordination accuracy of the array with each re-transmission. In a decentralized array, nodes only monitor the electrical states of their neighboring nodes and adjust their values to the average of its neighbors' and its own values. Because of this, the array topology is less strict, requiring only that all nodes have at least one edge connected to another node in the array. Centralized topologies are often simpler to implement and have the advantage of being able to synchronize the entire array in a single epoch (an atomic synchronization operation), while fully decentralized arrays are more involved and require multiple epochs to reach convergence, but are robust to any node entering and exiting the array at random. In this work, we discuss a centralized technique for simplicity, focusing on the electrical state estimation aspect of the coordination challenge; however, because this approach is fully digital (i.e., does not rely on any centralized analog reference hardware), a decentralized can easily be employed in practice. 

Once a timebase has been determined, the challenge of accurately comparing time, frequency, and phase offsets between elements within the array remains. In a spatially distributed network of \acp{sdr}, each radio will have its own free-running reference oscillator whose frequency of oscillation will fluctuate on many time scales due to varying intrinsic and extrinsic factors such as aging, temperature, pressure, and acceleration, to name a few.  Because the radios are distributed spatially---potentially on moving platforms---each radio will experience different extrinsic influences causing the oscillators to vary in an uncorrelated manner. Commonly, the task of phase estimation and correction is decomposed into three independent tasks, starting with either synchronization (time alignment) or syntonization (frequency alignment), which must be performed continuously, followed by static phase corrections to correct for the \ac{rfe} and antenna phase pattern, which can be performed prior to system operation. The task of time estimation can be accomplished using either one-way, or two-way techniques. One-way techniques have been implemented successfully in \acf{ntp} and \ac{gnss} time and frequency distribution; however, these techniques require the \ac{csi} and relative location and velocity of each node to be well-characterized to achieve high levels of accuracy, which is not always practical in a \ac{cda}. In situations where \ac{csi} is not available, \ac{twtt}-based techniques can be used in which the impact of the channel on internode time and phase measurements is implicitly cancelled by the two-way process, assuming it is quasi-static over the synchronization epoch~\cite{levine2008review,kirchner1991two-way}; a common implementation of this is IEEE-1588 \acf{ptp}, which can achieve synchronization to the microsecond-level in computer networks 
and has recently been extended to include a high-accuracy profile based on White Rabbit to achieve sub-nanosecond levels of synchronization~\cite{serrano2013white}. Several high-accuracy \ac{twtt}-based techniques have been experimentally demonstrated in the context of \ac{rf} wirelessly coordinated \acp{cda}, typically involving a multi-stage refinement technique to achieve \ac{toa} estimate accuracies significantly below the sampling period of the platform~\cite{pooler2018precise,prager2020wireless}. This technique is expanded upon in \cite{mghabghab2022adaptive-distri} and \cite{merlo2023tmtt} by taking advantage of the cooperative nature of the \ac{cda} system by leveraging spectrally-sparse two-tone \ac{toa} waveforms which minimize the variance on the \ac{toa} estimation while mitigating range-Doppler coupling when compared with the \ac{lfm} waveforms commonly used in \ac{cda} coordination. Frequency syntonization is commonly performed independently of time synchronization either directly via a continuous frequency reference broadcast and analog reception circuit, which is used to discipline the \acp{lo} of secondary nodes~\cite{abari2015airshare,chen2017wireless-synchr,mghabghab2021open-loop-distr,merlo2023tmtt,kiyaei2024interference-to}, or indirectly via digital spectral estimation of pulsed tones which are typically either estimated in the Fourier frequency domain digitally using peak estimation~\cite{quitin2013a-scalable-arch,werbunat2024synchronization-of-radars,kenney2024distributed-frequency-and-phase-synchronization,kenney2024frequency-synchronization}, or via carrier phase tracking over sequential pulses with sufficient periodicity to mitigate ambiguity~\cite{yan2019software-define,hanna2023distributed-transmit-beamforming-uav,holtom2024discobeam}. Similarly to time-transfer, these techniques can also be separated into one-way and two-way methods, with the latter being robust to Doppler shift, so long as the frequency shift is constant over a syntonization epoch~\cite{kenney2024distributed-frequency-and-phase-synchronization}.

In this work, we build on our previous research on high-accuracy wireless time coordination~\cite{merlo2023tmtt} by including several significant contributions. First, a new indirect digital frequency estimation technique which tracks the time drift of the reference oscillators is presented. This technique avoids the use of external frequency syntonization hardware making it amenable to implementation in commercially available software-defined radios, without modifications or additional hardware. Second, we present and experimentally verify a complete technique for the coordination of all fundamental electrical states (time, frequency, and phase) which is: resilient to multi-path and non-line-of-sight due to its reliance solely on the \ac{twtt} process~\cite{merlo2023ursigass}; resilient to relative motion and time-varying channel properties between nodes (assuming the channel is quasi-static over the synchronization epoch); Is both spectrally and temporally sparse, minimizing coordination overhead in both spectrum and time; and may be implemented using existing commercially available \acp{sdr} without requiring additional adjunct coordination hardware, which was required previously.
Finally, we provide a detailed treatment of sources of time, phase, and frequency error in the \ac{sdr} as well as detailed derivation of the electrical state relations.  

The new indirect frequency estimation technique described in this paper takes advantage of the fact that the \ac{rf} \ac{lo} on a typical \ac{sdr} platform will, on average, have the same fractional frequency error as the system clock which it is derived from. Because of this, if the frequency error of the system clock can be estimated with a time jitter significantly lower than the \ac{lo} period, the frequency error of the carrier can be indirectly estimated and compensated for.  This technique has several advantages over direct carrier spectral estimation and sequential carrier phase estimation techniques:
\begin{enumerate}
	\item \textit{Unambiguity:} The \ac{twtt} technique directly estimates the system clock's time instead of the carrier phase; thus, there is no ambiguity resolution required, unlike sequential carrier phase estimate-based techniques.
	\item \textit{Low Duty Cycle:} The variance of the \ac{twtt}-based frequency estimation is minimized by having short \ac{twtt} exchanges separated by a large amount of time, whereas the variance of spectral estimation techniques is minimized by integrating long duration frequency pulses. Because of this, the \ac{twtt}-based frequency estimation technique can provide lower variance estimates while maintaining low coordination overhead for the array.
 	\item \textit{Doppler and Multipath Tolerance:} The \ac{twtt}-based technique estimates the time independent of time-varying channel state from multipath and Doppler shift, due to its two-way nature, so long as the channel is quasi-static over the synchronization epoch. (i.e., the epoch is short relative to the evolution of the channel). While two-way direct spectral frequency estimates also separate the platform electrical states from the channel state, the requirement for increasing pulse duration to improve frequency estimation conflicts with the requirement to minimize the synchronization epoch duration to improve performance in time-varying channels.
\end{enumerate}
In \cite{merlo2023tmtt} it was shown that the high-accuracy \ac{twtt} technique could synchronize the system on the order of picoseconds, which should be sufficient to support beamforming directly at microwave frequencies based on prior analyses~\cite{nanzer2017open-loop-coher}. In this paper, we demonstrate this experimentally using the fully digital method proposed in this paper and compare it to the hybrid digital \ac{twtt} and analog wireless frequency syntonization technique described in~\cite{merlo2023tmtt}.

\section{Signal Model}
\label{sec:signal-model}

\begin{figure}
	\centering
	\includegraphics[width=1.0\columnwidth]{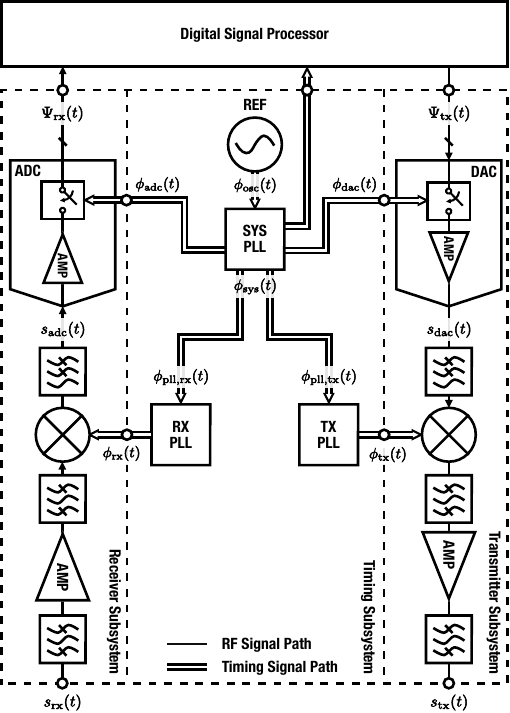}
	\caption{Simplified single channel full-duplex \acf{sdr} schematic used for system modeling. In general, all phase quantities are slowly time-varying with respect to changing extrinsic and intrinsic conditions such as device temperature, drive levels, and acceleration, among others.}
	\label{fig:sdr-diagram}
\end{figure}

In this work we consider an array of $N$ \acp{sdr} consisting of a typical direct conversion front end as shown in Fig. \ref{fig:sdr-diagram}.  The goal is for each of the \acp{sdr} to operate in phase coherence as if they shared the same frequency reference in order to achieve an active coherent array transmit gain of $N^2$; however, errors, primarily induced by the timing subsystem, will induce carrier frequency offset, sampling frequency offset (time scaling), and carrier and sampling phase wander.  In a distributed array, all these effects must be estimated and compensated for to ensure that the waveforms transmitted and received at each element are in phase at any given time, supporting beamforming operation.

\subsection{Intranode Signal Model}
\label{sec:intranode-signal-model}
From a signal perspective, the system may be broken into the transmitter, receiver, and timing subsystems, as shown in Fig. \ref{fig:sdr-diagram}. Following the signal from the perspective of the transmitter and receiver subsystems, each have four ports: the baseband signal at the data converters $\Psi_\mathrm{tx}(t),\Psi_\mathrm{rx}(t)$; the sample clocks $\phi_\mathrm{dac}(t),\phi_\mathrm{adc}(t)$; the \acp{lo} $\phi_\mathrm{tx}(t),\phi_\mathrm{rx}(t)$, which are often independently tunable; and the \ac{rf} ports $s_\mathrm{tx}(t),s_\mathrm{rx}(t)$. It should also be noted that a direct \ac{rf} sampling \ac{sdr} would behave in a similar manner, with the mixers being replaced by the data converters, using the \acp{lo} to drive the sample clocks, eliminating the phase uncertainty between the data converter clocks and \acp{lo}. To properly model the effect the system has on the transmitted and received waveforms, the impact of each subsystem must be modeled.

\subsubsection{Timing Subsystem}

The goal of the timing subsystem is to create a time and frequency reference which is true to an agreed upon frequency standard over the course of operation; however, practical challenges such as size, weight, power consumption and cost limit the options available for \acp{sdr}. Often, all of these parameters are minimized leading to the use of some form of crystal oscillators as the primary frequency reference in an \ac{sdr}, which typically have lower long-term stability and are more susceptible to environmental conditions than atomic frequency standards.  Environmental impacts such as temperature, pressure, humidity, and acceleration~\cite{allan1987time-and-freque,hellwig1990environmental-s} 
can all impact the frequency of oscillation, as well as initial manufacturing tolerance, crystal contamination creating loading, and internal crystalline structure changes, which can all contribute to aging frequency drift~\cite{walls1995fundamental-lim,
barnes1971characterizatio}. Aging is typically a slow process with fractional frequency drifts on the order of $1\times10^{-9}/$day for new crystals to $1\times10^{-12}/$day after several months and can be easily compensated by periodic recalibration over days~\cite{zhou2008frequency-accur,barnes1971characterizatio}. Environmental impacts on frequency are more challenging to mitigate. Typically, crystal oscillators are hermetically sealed to isolate them from contaminants, humidity, and pressure changes, and either temperature compensated (TCXO) or oven controlled (OCXO) by a heating element to keep the crystal at a constant temperature. While these minimize the impacts of environmental disturbances, they cannot be fully eliminated. Furthermore, on a moving platform, the impact of vibration and acceleration is very challenging to eliminate entirely, though various techniques have been implemented~\cite{kosinski2000theory-and-desi,mossammaparast2024techniques-on-c}.  Because the sources of environmental disturbances can be viewed as a random process, the impact of these sources on the oscillator phase can also be viewed as a random process inducing phase noise. Temperature and pressure changes typically induce phase noise very close to the carrier, while accelerations and vibration may produce spurs or phase noise at frequencies further from the carrier.  The resulting signal from the reference oscillator can be modeled as
\begin{align}
	\label{eq:oscillator-phase}
	\begin{split}
	\phi_\mathrm{osc}(t) &= 2\pi f_\mathrm{0,osc}\left[1 + \Delta f_\mathrm{osc}(t)\right]t + \nu_{\phi,\mathrm{osc}}(t) + \phi_{0,\mathrm{osc}}
	\end{split}
\end{align}
\begin{figure}
	\centering
	\includegraphics[width=0.9\columnwidth]{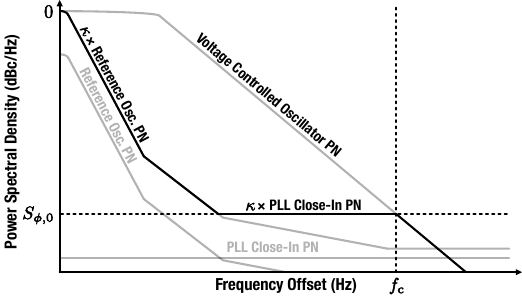}
	\caption{Phase noise profile of a \ac{pll} synthesizer, where $\kappa$ is the feedback divider ratio. The loop filter cutoff frequency is indicated by $f_c$ and $S_{\phi,0}$ is the noise floor of the \ac{pll} due to intrinsically generated noises.}
	\label{fig:pll-phase-noise}
\end{figure}%
where $\Delta f_\mathrm{osc}(t)=\delta f_\mathrm{0,osc}(t) / f_\mathrm{0,osc}$ is a fractional frequency error due to initial manufacturing defects and aging; $\delta f_\mathrm{0,osc}(t)$ is the absolute frequency error and the $f_\mathrm{0,osc}$ is the nominal frequency of the reference oscillator; and $\phi_\mathrm{0,osc}$ is the initial phase of the oscillator. The quantity $\nu_{\phi,\mathrm{osc}}(t)$ is the stationary zero mean random phase noise process which can be described by a gaussian power spectral density very near the carrier~\cite{
chorti2006a-spectral-mode} and a power law at further out frequencies~\cite{allan1987time-and-freque
},~\cite[\S 7.3]{gardner2005phaselock-techn}. 
This signal is then passed to the system \ac{pll} synthesizer (SYS PLL)---shown schematically in Fig. \ref{fig:pll-schematic}\footnote{Lettered figures, tables, and equations can be found in the appendices available in the supplemental materials on IEEE Xplore.}
---to generate a higher frequency reference for the \ac{dsp}, data converters, and the \ac{lo} \ac{pll} synthesizers.  In general, \acp{pll} act as a low-pass filter for phase noise from the reference source and the intrinsically generated phase noises due to resistive and active components, and a high-pass filter for \ac{vco} noise~\cite{gardner2005phaselock-techn}. An approximation of the phase noise profile at the output of a \ac{pll} synthesizer is shown in Fig. \ref{fig:pll-phase-noise}. The phase at the output of the SYS PLL can be modeled as the sum of the ideal output frequency and a time-varying phase error
\begin{align}
	\label{eq:sys-pll}
	\begin{split}
		\phi_\mathrm{sys}(t) &= 2\pi f_\mathrm{0,sys} t + \delta\phi_\mathrm{sys}(t)
	\end{split}
\end{align}
where
\begin{align}
	\label{eq:phase-error}
	\delta\phi_\mathrm{sys}(t) = 2\pi f_\mathrm{0,sys}\Delta f_\mathrm{osc} t + \nu_{\phi,\mathrm{sys}}(t) + \phi_\mathrm{0,sys}
\end{align}
and \mbox{$f_\mathrm{0,sys}=\kappa_\mathrm{sys} f_\mathrm{0,osc}$} where $\kappa_\mathrm{sys}$ is the feedback scaling factor for frequency multiplication, $\nu_{\phi,\mathrm{sys}}(t)=\kappa_\mathrm{sys}\langle\nu_{\phi,\mathrm{sys,in}}(t)+\nu_{\phi,\mathrm{osc}}(t)\rangle_\mathrm{lpf}+\langle\nu_{\phi,\mathrm{vco}}(t)\rangle_\mathrm{hpf}$ where $\nu_{\phi,\mathrm{sys,in}}(t)$ is the intrinsically generated phase noise, $\nu_{\phi,\mathrm{vco}}(t)$ is the phase noise generated by the \ac{vco}, and $\langle\cdot\rangle_\mathrm{lpf}$ and $\langle\cdot\rangle_\mathrm{hpf}$ are ideal low-pass and high-pass filters, respectively. The term $\phi_\mathrm{0,sys}$ is the static phase shift due to signal propagation delays and the initial random phase offset of the reference oscillator. 

The data converters and \ac{rf} \acp{pll} all behave in a similar manner; in addition, some \acp{dac} will include integrated \ac{pll} synthesizers and perform interpolation on the device to minimize analog filtering requirements. But in general, these can be viewed as adding a static phase delay and additional frequency scaling, while retaining the same fractional frequency stability as the reference oscillator. Here we will define the TX \ac{pll} \ac{lo} signal at the input of the mixer; the RX \ac{pll} \ac{lo} and data converter clocks $\phi_\mathrm{dac}(t)$ and $\phi_\mathrm{adc}(t)$ can be described similarly but are omitted for brevity. Starting with \eqref{eq:sys-pll} a second frequency scaling factor will be added $\kappa_\mathrm{tx}$; for the transmitter \ac{lo} letting $f_\mathrm{0,tx}=\kappa_\mathrm{sys}\kappa_\mathrm{tx} f_\mathrm{0,osc}$ yields
\begin{align}
	\label{eq:tx-lo}
	\begin{split}
	\phi_\mathrm{tx}(t) &= 2\pi f_\mathrm{0,tx}t + \delta\phi_\mathrm{tx}(t)
	\end{split}
\end{align}
where
\begin{align}
	\label{eq:tx-lo-error}
	\delta\phi_\mathrm{tx}(t) = 2\pi f_\mathrm{0,tx} \Delta f_\mathrm{osc} t + \nu_\mathrm{\phi,\mathrm{tx}}(t) + \phi_\mathrm{0,tx}
\end{align}
and $\nu_\mathrm{\phi,\mathrm{tx}}(t)$ is the summation of all phase noises at the \ac{lo} induced by the \acp{pll} as well as the phase errors from environmental perturbations and oscillator aging, and $\phi_\mathrm{0,tx}$ is the static phase offset due to propagation delay of the signal between the SYS PLL and TX PLL. The receiving \ac{lo} can be modeled in the same way. It's important to note that while $\nu_\mathrm{\phi,sys}(t)$ and $\nu_\mathrm{\phi,tx}(t)$ are represented as unique phase noise terms, these will be partially correlated noises deviating primarily at high offset frequencies from the carrier due to \ac{vco} noises.
 
For further analysis it is useful to consider the continuous representation of time \textit{in seconds} derived from a frequency reference, which is obtained from total accumulated phase divided by its nominal angular frequency
\begin{align}
	\label{eq:time-ref}
	T(t) = \frac{\phi(t)}{2\pi f_0} = t + \frac{\delta\phi(t)}{2\pi f_0}= t + \delta T(t).
\end{align}
From \eqref{eq:time-ref}, it's clear that our notion of \textit{time}---the accumulated phase at any given point in space---is relative to a given reference plane as well as the true angular rate of the phase $\mathrm{d}\,\phi(t)/\mathrm{d}t$; because the true value of frequency is not known, if the current angular rate of the phase is slower or faster than the expected frequency, the time will appear to run slower or faster relative to the global true time. In the case of system timing distribution, letting $\phi(t)\equiv\phi_\mathrm{sys}(t)$, \eqref{eq:time-ref} would become $T_\mathrm{sys}(t)$ representing the time at the output of the system frequency reference (SYS PLL); all components and transmission lines thereafter impart a phase shift on the timing signal path relative to this source.


\subsubsection{Transmitter Subsystem}

From the transmitter perspective, a baseband signal $\Psi_\mathrm{tx}(t)$ is generated in the \ac{dsp}, loaded into the \ac{dac} sample registers and shifted out at the rate of the sampling clock given by $f_\mathrm{s,dac}(t)=\mathrm{d}\,\phi_\mathrm{dac}(t)/ 2\pi\,\mathrm{d}t$ where $\phi_\mathrm{dac}(t)=\phi_\mathrm{sys}(t-\tau_\mathrm{dac})$ is the \ac{dac} sampler clock phase and $\tau_\mathrm{dac}$ is the propagation delay for the signal to travel from the SYS PLL output to the clocking circuit on the \ac{dac}; because the clock signal is periodic, a time delay is indistinguishable from a phase shift, thus $\tau_\mathrm{dac}$ can be grouped with the static phase term $\phi_\mathrm{0,dac}$. The waveform at the output of the sampler can be modeled as a pulse train of sampling impulses, whose bandwidths are determined by the analog bandwidth of the sampler and output filters $\beta_\mathrm{dac}$, with a sampling period $\tau_\mathrm{s,dac}=1/f_\mathrm{s,dac}$; for simplicity, we will assume that $f_\mathrm{s,dac}=f_\mathrm{0,sys}=1/\tau_\mathrm{s,sys}$, yielding a reconstructed signal modeled by
\begin{align}
	\begin{split}
		s_\mathrm{dac}(t) &= \begin{multlined}[t][\columnwidth-\widthof{$s_\mathrm{dac}(t) = $ }]
 		\sum_k \Psi_\mathrm{tx}(k\tau_\mathrm{s,sys})\sinc\left\{\beta_\mathrm{dac}[t-T_\mathrm{dac}(k\tau_\mathrm{s,sys})]\right\}
		\end{multlined}
	\end{split}
\end{align}
where $\sinc(x)=\sin(\pi x) / \pi x$ and $T_\mathrm{dac}(t)$ is the time at the \ac{dac} derived from $\phi_\mathrm{dac}(t)$ using \eqref{eq:time-ref}. Assuming the drive level of the \ac{dac} amplifier is well below its compression point, third order harmonics should be negligible.
The upconverted signal after mixing with the TX \ac{lo} given by \eqref{eq:tx-lo} 
is
\begin{align}
	\label{eq:s-tx}
	\begin{split}
	s_\mathrm{tx}(t) &= \begin{multlined}[t][\columnwidth-\widthof{$s_\mathrm{tx}(t) = $ }]
		\exp\{j\left[2\pi f_\mathrm{0,tx}t + \delta\phi_\mathrm{tx}(t)\right]\}\sum_k \Psi_\mathrm{tx}(k\tau_\mathrm{s,sys})\\
		\!\!\!\times\sinc\bigg\{\beta_\mathrm{dac}\bigg[t-\tau_\mathrm{tx}-k\tau_\mathrm{s,sys}-\delta T_\mathrm{dac}(k\tau_\mathrm{s,sys})\bigg]\bigg\}
	\end{multlined}
	\end{split}
\end{align}
where $\tau_\mathrm{tx}$ is the propagation delay of the signal measured between the \ac{dac} and the \ac{rf} port, and again assuming the amplifier is well below its compression point and proper filtering is applied to mitigate mixing images. 

From \eqref{eq:s-tx} it is shown that the clock frequency drift and phase noise will have an effect on both the sampler and the mixer. The impact on the sampler is time scaling due to the mismatch between expected sampling rate and actual sampling rate, creating a sampling frequency shift proportional to $\Delta f_\mathrm{osc} f_\mathrm{0,sys}$ which manifests as an integrated sampling time error; in this case, if the clock is too fast relative to the agreed upon standard of the array, a radio will transmit its message faster, meaning it will be a shorter duration and its entire frequency spectra will be scaled by $1+\Delta f_\mathrm{osc}$ meaning it will have greater bandwidth. Additionally, a static sampling offset due to initial clock phases $\phi_\mathrm{0,dac}$ and device startup time is also present. The impact of the time and frequency errors on the output of the mixer is that of a phase shift and a frequency offset.  The frequency error will shift the entire baseband by the error amount and the static phase offset will directly result in a phase offset at the carrier. From \eqref{eq:s-tx}, it is shown that these combined effects result in a net transmit center frequency of
\begin{align}
	\label{eq:tx-freq}
	f_\mathrm{tx}=(1 + \Delta f_\mathrm{osc})\left(f_\mathrm{0,if}+f_\mathrm{0,tx}\right)
\end{align}
where $f_\mathrm{0,if}$ is any digitally generated \ac{if}.


\subsubsection{Receiver Subsystem}
The signal received at the \ac{adc} can be modeled similarly
\begin{align}
	\label{eq:s-adc}
\begin{split}
\Psi_\mathrm{rx}[k]	&= \!\begin{multlined}[t][\columnwidth-\widthof{$s_\mathrm{adc}[k] =$}]
	\exp\{-j[2\pi f_\mathrm{0,rx} t + \delta\phi_\mathrm{rx}(t)]\}\\
	\times s_\mathrm{rx}(t-\tau_\mathrm{rx}) \,\delta\!\left[t-k\tau_\mathrm{s,sys}-\delta T_\mathrm{adc}(k\tau_\mathrm{s,sys})\right]	
\end{multlined}
\end{split}
\end{align}
where $\delta(x)$ is the Dirac delta function representing an ideal sampling operation.

The clock frequency and phase errors will result in a similar error on the receiver with one important difference: the time and frequency errors will manifest in the opposite direction. Continuing with the example of reference oscillator running faster than the agreed upon frequency standard, this will cause the receive signal to be sampled at a higher rate than expected and thus sample a smaller period of the incoming waveform than expected.  This will thus have the result of scaling the entire spectra in the \textit{opposite} direction of the transmitter making it appear to scale in bandwidth by a factor of $1-\Delta f_\mathrm{osc}$. Similarly to the transmitter, this still results in an integrated sample time error and an initial sampling time offset due to the static sampling clock phase offsets. The impact of frequency and phase error on the mixer will also create the opposite effect of the transmitter, shifting the received signal down by a greater amount than expected and inducing a static phase rotation in the opposite direction from the transmitter, which can be seen by the negative in the exponential term of \eqref{eq:s-adc}. The combined effect on the received frequency of a single tone $f_\mathrm{rx}$ after downconversion and sampling at the receiver is thus
\begin{align}
	\label{eq:rx-freq}
	\begin{split}
	f_\mathrm{\Psi,rx}&=\frac{1}{\Delta f_\mathrm{osc}}\left(f_\mathrm{rx}-\Delta f_\mathrm{osc}f_\mathrm{0,lo}\right)-f_\mathrm{0,if}\\
	&=\frac{f_\mathrm{rx}}{\Delta f_\mathrm{osc}}-f_\mathrm{0,lo}-f_\mathrm{0,if}
\end{split}
\end{align}
where the received frequency is scaled by the inverse of the fractional reference oscillator error and translated down in frequency by the \ac{rf} \ac{lo} and digital \ac{if}.

\subsection{Internode Signal Model}
\label{sec:internode-signal-model}

Because the goal is to determine the error between nodes in the array and not relative to an external standard, we are primarily interested in estimating the differences in the electrical states between any two nodes in the array. 
To accomplish this, the phase of the information and the carrier must be continuously aligned between nodes; the carrier phase difference can be represented as\footnote{Full derivations for \eqref{eq:lo-error}, \eqref{eq:apparent-tof}--\eqref{eq:tof-approx}, and \eqref{eq:freq-error} can be found in Appendix \ref{sec:derivations} in the supplemental materials on IEEE Xplore.}
\begin{align}
\begin{split}
	\label{eq:lo-error}
	\phi_\mathrm{lo}^{(n,m)}(t)&=\phi_\mathrm{lo}^{(m)}(t) - \phi_\mathrm{lo}^{(n)}(t)
	= \delta\phi_\mathrm{lo}^{(m)}(t) - \delta\phi_\mathrm{lo}^{(n)}(t)\\
	&=\!\begin{multlined}[t]
	2\pi f_\mathrm{0,lo} \Delta f_\mathrm{osc}^{(n,m)} t + \phi_\mathrm{0,lo}^{(n,m)} + \nu^{(n,m)}_\mathrm{\phi,\mathrm{lo}}(t)
	\end{multlined}
\end{split}
\end{align}
where $\phi_\mathrm{lo}^{(n)}(t)$ is the \ac{lo} phase at the $n$th node and, for simplicity, we will assume that $\phi_\mathrm{lo}^{(n)}(t)=\phi_\mathrm{tx}^{(n)}(t)=\phi_\mathrm{rx}^{(n)}(t)$ meaning the \acp{lo} are tuned to the same frequency, and $\nu^{(n,m)}_\mathrm{\phi,\mathrm{lo}}(t)\equiv\sqrt{2}\nu_\mathrm{0,lo}^{(n)}$, assuming $\nu_\mathrm{0,lo}^{(n)}$ and $\nu_\mathrm{0,lo}^{(m)}$ have the same \ac{psd}, due to incoherent summation. The quantity $\Delta f_\mathrm{osc}^{(n,m)}$ is the difference in relative frequency errors between nodes $n$ and $m$. 
The system \ac{pll} time difference between nodes can be found in a similar manner to the \ac{lo} phase by using the relation defined in \eqref{eq:time-ref}
\begin{align}
\label{eq:internode-time-error}
\begin{split}
	T_\mathrm{sys}^{(n,m)}(t) &= T_\mathrm{sys}^{(m)}(t) - T_\mathrm{sys}^{(n)}(t)\\
	&=\!\begin{multlined}[t]
	\Delta f_\mathrm{osc}^{(n,m)} t + T_\mathrm{0,sys}^{(n,m)}
	+ \frac{\nu^{(n,m)}_\mathrm{\phi,\mathrm{sys}}(t)}{2\pi f_\mathrm{0,sys}}
	\end{multlined}
\end{split}
\end{align}
where the $\delta\phi_\mathrm{sys}(t)$ terms are expanded using \eqref{eq:phase-error}.

It is also useful to analyze this error as a function of frequency in the presence of relative motion between platforms. Using \eqref{eq:tx-freq} as the signal transmitted by node $N_n$, the received waveform at node $N_m$ can be modeled by
\begin{align}
	f_\mathrm{rx}^{(m)} = \frac{v_r^{(m,n)}}{\lambda^{(n)}}+f_\mathrm{tx}^{(n)}=f_\mathrm{tx}^{(n)}\left(1+\frac{v_r^{(m,n)}}{c}\right)
\end{align}
where $v_r^{(m,n)}=\mathrm{d}R^{(m,n)}(t)/\mathrm{d}t$ is the radial closing velocity of node $N_n$ relative to node $N_m$, $\lambda^{(n)}=c/f_\mathrm{tx}^{(n)}$ and $c$ is the speed of light in the medium; $R^{(m,n)}(t)$ is the radial distance between nodes $n$ and $m$. The \textit{apparent} received frequency after downconversion and sampling at node $N_m$ can then be modeled using \eqref{eq:tx-freq} and \eqref{eq:rx-freq} as
\begin{align}
	\label{eq:apparent-doppler}
	\begin{split}
		\bar{f}_d^{(n,m)} &=f_\mathrm{0,rf}\frac{\Delta f_\mathrm{osc}^{(n)}}{\Delta f_\mathrm{osc}^{(m)}}\left(1+\frac{v_r^{(m,n)}}{c}\right)-f_\mathrm{0,rf}
	\end{split}
\end{align}
where $f_\mathrm{0,rf}=f_\mathrm{0,if}+f_\mathrm{0,lo}$. We will refer to this quantity as the \textit{apparent Doppler shift} because it manifests as a frequency shift indistinguishable from the true Doppler shift, but is caused by both the frequency error between nodes and the true Doppler shift.


\section{Electrical State Estimation}
\label{sec:electrical-state-estimation}

\subsection{Time and Phase Estimation}
\label{sec:time-estimation}
\begin{figure}
	\centering
	\includegraphics[width=\columnwidth]{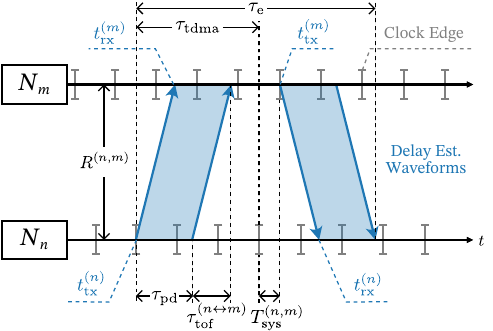}
	\caption{\Acf{twtt} diagram using a \acf{tdma} scheme showing a single synchronization epoch. In this channel multiplexing scheme, each node has a scheduled transmission time separated by $\tau_\mathrm{tdma}$. The internode range $R^{(n,m)}$ and relative time error $\delta T_\mathrm{sys}^{(n,m)}$ are shown as quasi-static over the duration of the epoch; thus, $\tau_\mathrm{tof}^{(n\leftrightarrow m)}$ is also reciprocal and quasi-static during the epoch. Due to synchronization errors, node $N_m$ transmits its message slightly misaligned with the start of its \ac{tdma} frame.}
	\label{fig:twtt}
\end{figure}

\begin{figure}
	\centering
	\includegraphics[width=\columnwidth]{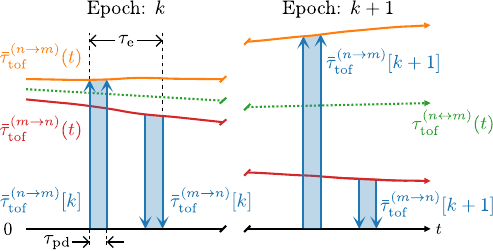}
	\caption{Apparent \acf{tof} diagram for the \ac{twtt} process. The nonlinearities in the $\bar{\tau}_\mathrm{tof}$ traces are magnified for illustration purposes, but care should be taken to ensure that $\tau_\mathrm{e}$ is short enough that the apparent \ac{tof} between nodes is quasi-static during the synchronization epoch duration to ensure the assumption of channel reciprocity during the \ac{twtt} exchange is not violated.}
	\label{fig:tof}
\end{figure}

In this work we utilize the high-accuracy \ac{twtt} methods described in~\cite{merlo2023tmtt}, shown graphically in Figs. \ref{fig:twtt} and \ref{fig:tof}. This process uses a traditional \ac{twtt} process to estimate the time offsets between the system clocks on each platform~\cite{levine2008review,hanson1989fundamentals,kirchner1991two-way}; however, this process is limited to the accuracy of the time of transmission and arrival estimates from each \ac{sdr}. On the transmit side, the radio transmit time can be scheduled to align with a clock edge relative to the system time; the \ac{dac} and \ac{rfe} do not appreciably increase the jitter, so the time of transmission can often be estimated with high certainty. Estimation of the \ac{toa} is more challenging because the received waveform can arrive at any point between clock edges; thus, waveform optimization and \ac{toa} refinement based on ~\cite{merlo2023tmtt} is performed to achieve high-accuracy time estimation using two-tone waveforms by leveraging the cooperative nature of the system.

\subsubsection{Two-Way Time Transfer}
\label{sec:twtt}

The goal of the \ac{twtt} process is to estimate the accumulated clock phase error between the reference oscillators at each \ac{sdr}.  Using \eqref{eq:tx-lo} and \eqref{eq:time-ref} we can define the time at the data converters as $T_\mathrm{dac}(t)$ and $T_\mathrm{adc}(t)$ from which we can define the \textit{apparent \acf{tof}} between nodes $N_n$ and $N_m$ during the $k$th synchronization epoch
\begin{align}
	\label{eq:apparent-tof}
	\begin{split}
		\bar{\tau}_\mathrm{tof}^{(n\rightarrow m)}[k] &= T_\mathrm{adc}^{(m)}\!\left(t_\mathrm{rx}^{(m)}[k]\right) - T_\mathrm{dac}^{(n)}\!\left(t_\mathrm{tx}^{(n)}[k]\right)\\
		&= T_\mathrm{sys}^{(m)}\!\left(t_\mathrm{tx}^{(n)}[k]+\tau_\mathrm{tof}^{(n\rightarrow m)}[k]\right) - T_\mathrm{sys}^{(n)}\!\left(t_\mathrm{tx}^{(n)}[k]\right)\\
		 &= \tau_\mathrm{tof}^{(n\rightarrow m)}[k]\left(1+\Delta f_\mathrm{osc}^{(m)}\right)+ T_\mathrm{0,sys}^{(n,m)}\\
		&\quad+\Delta f_\mathrm{osc}^{(n,m)}t_\mathrm{tx}^{(n)}[k]+\frac{\nu_\mathrm{\phi,dac}[k]+\nu_\mathrm{\phi,adc}[k]}{2\pi f_\mathrm{0,sys}}
	\end{split}
\end{align}
where $t_\mathrm{tx}^{(n)}[k]$ and $t_\mathrm{rx}^{(m)}[k]$ are the actual times that the waveforms were transmitted and received at nodes $N_n$ and $N_m$, respectively, and 
$\tau_\mathrm{tof}^{(n\rightarrow m)}[k]$ is the true \ac{tof} between nodes $N_n$ and $N_m$. We will assume a static phase calibration has been performed at each radio eliminating the static time errors at the data converters $\left(\text{i.e., } T_\mathrm{0,dac}^{(n,m)}= T_\mathrm{0,adc}^{(n,m)}=0\right)$, which allows simplification to only terms of $T_\mathrm{sys}(t)$ at each node.
The quantity $\bar{\tau}_\mathrm{tof}^{(n\rightarrow m)}[k]$ is referred to as the \textit{apparent \ac{tof}} because the time estimate is derived from the clock estimates including local clock error; the $\rightarrow$ notation in the superscript indicates that this quantity may not be reciprocal. The time offsets between nodes can then be estimated by
\begin{align}
	\label{eq:twtt-full}
	\begin{split}
		\hat{T}_\mathrm{}^{(n,m)}[k] &= \frac{\bar{\tau}_\mathrm{tof}^{(n\rightarrow m)}[k] - \bar{\tau}_\mathrm{tof}^{(m\rightarrow n)}[k]}{2}\\
		&\approx T_\mathrm{sys}^{(n,m)}[k]+\frac{1}{2}\Delta f_\mathrm{osc}^{(n,m)}\left(\tau_\mathrm{tof}^{(n\leftrightarrow m)}+\tau_\mathrm{tdma}\right)\\
		 &\quad+\nu_\mathrm{\phi}[k].
	\end{split}
\end{align}
where $\nu_\mathrm{\phi}[k]$ is the combined averaged phase noise term and 
$\tau_\mathrm{tof}^{(n\leftrightarrow m)}=\tau_\mathrm{tof}^{(n\rightarrow m)}=\tau_\mathrm{tof}^{(m\rightarrow n)}$ is the bidirectional \ac{tof}, assuming the \ac{tof} is quasi-static over the synchronization epoch $\tau_\mathrm{e}=\tau_\mathrm{tof}^{(n\leftrightarrow m)} + t_\mathrm{tx}^{(n,m)} + \tau_\mathrm{pd}$, where $\tau_\mathrm{pd}$ is the \ac{twtt} message pulse duration. The approximation in \eqref{eq:twtt-full} can be made once the system times are closely aligned, $t_\mathrm{tx}^{(m)}\approx t_\mathrm{tx}^{(n)}+\tau_\mathrm{tdma}$, where $\tau_\mathrm{tdma}$ is the \ac{tdma} time slot duration (see Fig. \ref{fig:twtt}). 
Furthermore, if the product of $\Delta f_\mathrm{osc}^{(n,m)}\tau_\mathrm{tof}^{(n\leftrightarrow m)}$ is small relative to the level of accuracy required, it may be neglected (e.g., for oscillators with $\Delta f_\mathrm{osc}^{(n,m)}=6$\,ppm over a 100\,m channel, the error would be $\approx\,$1\,ps).  Similarly, if the \ac{tdma} period is kept short and the product of $\Delta f_\mathrm{osc}^{(n,m)}\tau_\mathrm{tdma}$ is small, the impact of clock drift over the \ac{tdma} window can also be neglected (e.g., for oscillators with $\Delta f_\mathrm{osc}^{(n,m)}=6$\,ppm with $\tau_\mathrm{tdma}=\SI{10}{\micro\second}$, the error would be $\approx\,$30\,ps); acceptable levels of accuracy will be determined by the application. With these assumptions \eqref{eq:twtt-full} becomes
\begin{align}
	\label{eq:twtt-approx}
	\begin{split}
		\hat{T}^{(n,m)}[k] &\approx T_\mathrm{sys}^{(n,m)}[k] +\nu_\mathrm{\phi}[k]
	\end{split}
\end{align}
where $T_\mathrm{sys}^{(n,m)}$ is given by \eqref{eq:internode-time-error}.
From \eqref{eq:twtt-full} it can also be seen that the time offset estimate does not depend on the path that the signal traversed between radios, meaning that the technique is robust to \ac{nlos} scenarios, so long as the path taken is reciprocal~\cite{merlo2023ursigass}.
In a similar way, the internode \ac{tof} can be estimated by simply taking the average of the apparent \acp{tof}
\begin{align}
	\label{eq:tof-full}
	\begin{split}
		\hat{\tau}_\mathrm{tof}^{(n\leftrightarrow m)}[k] &= \frac{\bar{\tau}_\mathrm{tof}^{(n\rightarrow m)}[k] + \bar{\tau}_\mathrm{tof}^{(m\rightarrow n)}[k]}{2}\\
		&=\tau_\mathrm{tof}^{(n\leftrightarrow m)}+\frac{1}{2}\tau_\mathrm{tof}^{(n\leftrightarrow m)}\left(\Delta f_\mathrm{osc}^{(n)}+\Delta f_\mathrm{osc}^{(m)}\right)\\
		&\qquad-\frac{1}{2}\Delta f_\mathrm{osc}^{(n,m)}\tau_\mathrm{tdma}+\nu_\mathrm{\phi}[k]
	\end{split}
\end{align}
which, similar to the time offset estimation, the oscillator error term may be neglected if the product of the \ac{tof} and average oscillator error is small or well synchronized (e.g., for oscillators with $\Delta f_\mathrm{osc}^{(n,m)}=6$\,ppm over a 100\,m channel, the error would be $\approx\,$1\,ps, or \SI{300}{\micro\meter}). The \ac{tdma} period may also be neglected if the product of $\Delta f_\mathrm{osc}^{(n,m)}\tau_\mathrm{tdma}$ is small (e.g., for oscillators with $\Delta f_\mathrm{osc}^{(n,m)}=6$\,ppm with $\tau_\mathrm{tdma}=\SI{10}{\micro\second}$, the error would be $\approx\,-$30\,ps, or \SI{-9}{\milli\meter})
\begin{align}
	\label{eq:tof-approx}
	\begin{split}
		\hat{\tau}_\mathrm{tof}^{(n\leftrightarrow m)}[k] \approx \tau_\mathrm{tof}^{(n\leftrightarrow m)}+\nu_\mathrm{\phi}[k].
	\end{split}
\end{align}
From the \ac{tof} estimate, the internode range can be estimated directly by $\hat{R}^{(n\leftrightarrow m)}[k]=c\cdot\hat{\tau}_\mathrm{tof}^{(n\leftrightarrow m)}[k]$. It is worth noting here that \eqref{eq:tof-full} shows that accuracy depends not only on the relative internode frequency error as in \eqref{eq:twtt-full}, but also the mean frequency error of the system $\left(\Delta f_\mathrm{osc}^{(n)} + \Delta f_\mathrm{osc}^{(m)}\right)/2$. This is because the \ac{tof} is relative to the frequency standard used by the system, i.e., if range in meters is desired, the timing deviation from the \acf{si} definition of the second will result in additional ranging error.  It is also important to note that the \ac{tof} is not necessarily the shortest path between elements---typically the quantity that is desired for beamforming phase weighting computations---as it may be impacted by multipath or \ac{nlos}. To an extent this may be mitigated in multipath scenarios by choosing a waveform with high occupied bandwidth to resolve multipath scattering; however, this is not always feasible and, as will be discussed in Section \ref{sec:high-accuracy-toa} increases the lower-bound on \ac{toa} estimation variance relative to a sparse waveform with low occupied bandwidth (i.e., low spectral occupancy relative to the maximum spectral extent of the waveform).

\subsubsection{High-Accuracy Time of Arrival Estimation}
\label{sec:high-accuracy-toa}

To achieve high-accuracy \ac{toa} estimation, the waveform is first optimized to minimize the time estimation variance lower bound for a given bandwidth by studying the \ac{crlb} for \ac{toa} estimation, then a multi-step time estimation refinement process is implemented to estimate the delay to below a single clock tick on the \ac{adc} using matched filtering, \ac{qls} peak refinement, and a \ac{lut} to remove residual bias due to \ac{qls} by leveraging the cooperative nature of the system.

In the \ac{twtt} exchange, a narrowband received signal at the \ac{rf} port $s_\mathrm{rx}^{(n)}(t)$ can be modeled as a time delayed and frequency shifted copy of $s_\mathrm{tx}^{(m)}(t)$ with added noise
\begin{align}
	s_\mathrm{adc}^{(m)}(t)=\alpha s_\mathrm{tx}^{(n)}\!\left(t-\bar{\tau}_\mathrm{tof}^{(n\rightarrow m)}\right) \exp{\left(j 2\pi \bar{f}_d^{(n,m)} t\right)} + w(t)
\end{align}
where $\alpha$ is a complex channel weight describing the phase shift due to propagation delay and amplitude scaling due to channel loss and $w(t)$ is a white Gaussian noise process with a \ac{psd} of $\nu_0/2$ due to timing phase noises and thermal noise.
The lower bound on apparent time delay and apparent Doppler shift can then be found by solving the \ac{crlb} inequality with respect to each parameter resulting in~\cite[\S6.3]{van-trees2004detection-estim}\cite{nanzer2016bandpass-signal,nanzer2017accuracy}
\begin{align}
	\label{eq:var-tof}
	\mathrm{var}\!\left(\hat{\bar{\tau}}^{(n\rightarrow m)}_\mathrm{tof}-\bar{\tau}^{(n\rightarrow m)}_\mathrm{tof}\right)\ge \frac{\nu_0}{2|\alpha|^2}\left(\zeta_f^2-\frac{1}{E_s}\mu_f^2 \right)^{-1}
\end{align}
where $|\alpha|^2/\nu_0$ is the post-integration \ac{snr}, $E_s \triangleq \int |s_\mathrm{tx}(t)|^2\,\mathrm{d}t$ is the total signal energy, $\zeta_f^{2}\triangleq \int (2\pi f)^2|S_\mathrm{tx}(f)|^2\,\mathrm{d}f$ is the mean-square bandwidth, and $\mu_f \triangleq  \int 2\pi f |S_\mathrm{tx}(f)|^2\,\mathrm{d}f $ is the mean frequency. The apparent Doppler shift is
\begin{align}
	\label{eq:var-dop}
	\mathrm{var}\!\left(\hat{\bar{f}}^{(n,m)}_d-\bar{f}^{(n,m)}_d\right)\ge \frac{\nu_0}{2|\alpha|^2}\left(\zeta_t^2-\frac{1}{E_s}\mu_t^2 \right)^{-1}
\end{align}
where $\zeta_t^{2}\triangleq \int (2\pi t)^2|s_\mathrm{tx}(t)|^2\,\mathrm{d}t$ is the mean-square duration, and $\mu_t^{2}\triangleq\int 2\pi t |s_\mathrm{tx}(t)|^2\,\mathrm{d}t $ is the mean time. For signals centered around dc, the mean frequency can be omitted; similarly, by choosing a time reference at the center of the waveform, the mean time may also be omitted. 
The two-way accuracy bound for both the time error and \ac{tof} may then be obtained simply by taking the average of the variances of the one-way \ac{tof} estimate
\begin{align}
	\label{eq:crlb-twtt}
	\begin{multlined}[\columnwidth]
		\mathrm{var}\!\left( \hat{T}_\mathrm{}^{(n,m)}- {T}_\mathrm{}^{(n,m)}\right) \\
		\ge \frac{1}{2} \bigg[ \mathrm{var}\!\left(\hat{\bar{\tau}}^{(n\rightarrow m)}_\mathrm{tof}-\bar{\tau}^{(n\rightarrow m)}_\mathrm{tof}\right) 
		+ \mathrm{var}\!\left(\hat{\bar{\tau}}^{(m\rightarrow n)}_\mathrm{tof}-\bar{\tau}^{(m\rightarrow n)}_\mathrm{tof}\right) \bigg]
	\end{multlined}
\end{align}
which implies, given identical waveforms in both directions of the \ac{twtt} exchange, the two-way accuracy is inversely proportional to the average \ac{snr} seen at each receiver.

From \eqref{eq:var-tof} and \eqref{eq:var-dop} it is evident that lower bound on variance of both time and frequency estimation is inversely proportional to \ac{snr} at the receiver. However, in \eqref{eq:var-tof} the signal is also proportional to the inverse of the mean-square bandwidth $\zeta_f^2$ which is maximized by moving all of the energy of the frequency spectrum to the edges of the available bandwidth, e.g., two infinite duration tones separated by maximum available bandwidth; in \eqref{eq:var-dop} the signal is also inversely proportional the mean-square duration $\zeta_t^2$ which is maximized by placing all of the signal energy at the edges of the available pulse duration, e.g., two infinite bandwidth pulses at the edges of the pulse duration. This leads to conflicting requirements, thus making it impossible to simultaneously minimize both instantaneous \ac{toa} estimation and instantaneous frequency offset estimation. In this work, we choose to achieve the highest available instantaneous estimation of the relative clock phase error between systems (fast-time) and derive the frequency error from sequential estimates of time (slow-time); we utilize a \ac{ptt} waveform which approximates placing the signal energy at the edges of the available bandwidth in fast-time, and over long time intervals we will obtain relatively short duration pulses separated by a large duration of time, approximating the optimal frequency estimation technique of widely separated impulses in time.

\begin{figure}
	\centering
	\includegraphics[width=\columnwidth]{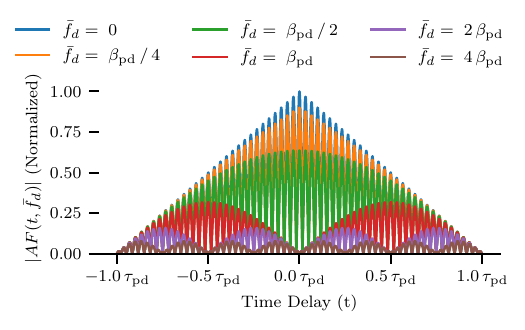}
	\caption{Ambiguity function slices of constant apparent Doppler shift $\bar{f}_d$ given by \eqref{eq:ambiguity-function} with \acf{ptt} separation \hbox{$\beta_\mathrm{ptt}=30/\tau_\mathrm{pd}$} and pulse bandwidth $\beta_\mathrm{pd}=1/\tau_\mathrm{pd}$. Several slices are plotted showing that as $\bar{f}_d$ approaches $\beta_\mathrm{pd}/2$, the peak of the magnitude of the matched filter no longer corresponds to the zero-delay lobe meaning a simple peak find can no longer be used.}
	\label{fig:ambiguity-function}
\end{figure}

The \ac{ptt} waveform is ideally described by
\begin{align}
	\label{eq:two-tone}
	s(t) = \Pi\!\left(\frac{t}{\tau_\mathrm{pd}}\right)\left(e^{-j\pi \beta_\mathrm{ptt} t}+e^{j\pi \beta_\mathrm{ptt} t}\right)
\end{align}
where $\Pi(t/\tau_\mathrm{pd})$ is a rectangular windowing function with a duration $\tau_\mathrm{pd}$ and $\beta_\mathrm{ptt}=f_2-f_1$ is the tone separation, assumed to be centered about dc after downconversion and sampling. The half-power bandwidth of each tone due to the rectangular pulse duration is $\beta_\mathrm{pd} = 1/ \tau_\mathrm{pd}$. For a \ac{ptt} waveform $\zeta_\mathrm{f(ptt)}=\pi\beta_\mathrm{ptt}$ and $\zeta_\mathrm{t(ptt)}=\pi\tau_\mathrm{pd}/\sqrt{3}$~\cite{nanzer2017accuracy}.

To perform \ac{toa} estimation a matched filter is formulated which maximizes the \ac{snr} in the presence of additive white gaussian noise, which is simply the convolution of the received waveform with the complex conjugate of the transmitted waveform~\cite[\S4.2]{richards2014fundamentals}. However, the impact of frequency error between the matched filter and the received waveform causes a distortion in the ideal matched filter. The impact of time and frequency offsets due to $\bar{\tau}_\mathrm{tof}^{(n\rightarrow m)}$ and $\bar{f}_{d}^{(n, m)}$ on the matched filter is given by its ambiguity function~\cite{ellison2020high-accuracy-w}
\begin{align}
	\label{eq:ambiguity-function}
	\begin{split}
		&\left|\mathrm{AF}(t, \bar{f}_d)\right| = \left|\int_{-\infty}^\infty s^*(\tau-t)s(\tau)e^{j2\pi \bar{f}_d \tau} d\tau\right|\\
		& \begin{multlined}[\columnwidth] =
		\Bigg| \frac{e^{j\pi \bar{f}_d t}}{4} \Lambda\!\left(\frac{t}{\tau_\mathrm{pd}}\right) \Bigg\{ \left(e^{-j\pi \beta_\mathrm{ptt} t}+e^{j\pi \beta_\mathrm{ptt} t}\right) \\
		\times \sinc\left[  \bar{f}_d \Lambda\!\left(\frac{t}{\tau_\mathrm{pd}}\right) \right] 
		+ \sinc\!\left[  (\beta_\mathrm{ptt}+\bar{f}_d) \Lambda\!\left(\frac{t}{\tau_\mathrm{pd}}\right) \right] \\
		\qquad \qquad + \sinc\!\left[  (\bar{f}_d-\beta_\mathrm{ptt}) \Lambda\!\left(\frac{t}{\tau_\mathrm{pd}}\right) \right] \Bigg\} \Bigg|
		\end{multlined}
	\end{split}
\end{align}
 where $\Lambda\!\left(t/\tau_\mathrm{pd}\right)$ is a triangular amplitude windowing function of duration $2\tau_\mathrm{pd}$.  Slices of the normalized ambiguity function are plotted in Fig. \ref{fig:ambiguity-function} using the default parameters used in the experiments in this work.  By looking at the zero-Doppler cut $\bar{f}_d=0$ of the ambiguity function, the periodic structure of the matched filter response to the \ac{ptt} waveform can be seen which includes a triangular envelope due to the convolution of the rectangular pulse envelope modulated by a periodic structure whose periodicity is equal to the inverse of the beat frequency between tone carriers $1/\beta_\mathrm{ptt}$; due to this fact, the \ac{ptt} waveform is not used for radar ranging as the sidelobes would significantly degrade image quality in any complex scene; however, in a cooperative two-way system, this waveform can be used so long as the pulse waveforms do not overlap in time, frequency, and space. Along the zero-delay cut $t=0$ a sinc shape is produced due to the rectangular pulse modulation envelope in the frequency domain. Of significant interest is the fact that the triangular shape of the matched filter begins to flatten out under moderate apparent Doppler shifts, meaning that the ability to pick the correct peak using a simple peak-finding operation decreases under noise; furthermore, at a certain frequency error, the peak of the matched filter no longer represents the true time delay at all. This cross-over point appears at $\bar{f}_d=1/(2\tau_\mathrm{pd})$, meaning a simple peak find will no longer yield the correct delay if the apparent Doppler shift is greater than half the pulse bandwidth, shown in green in Fig. \ref{fig:ambiguity-function}; practically, a larger margin will be required due to added noise which can cause amplitude fluctuations in the $1/\beta_\mathrm{ptt}$ lobes of the matched filter. Generally, this implies that shorter pulses will be more resilient to frequency offset and Doppler shift. This conflicts with the \ac{crlb} for \ac{toa} estimation described in \eqref{eq:var-tof}; thus, a balance in \ac{ptt} duration must be made between minimizing time estimation ability and resilience to frequency offset and Doppler shift which will be dependent on the application.
 
While the matched filtering process produces a waveform at its output that, under small to moderate apparent Doppler shifts, maximizes the output power at the true time delay of the waveform, it is still discretized to the sampling period of the platform which, for non-\ac{rf} sampling \acp{sdr}, is significantly larger than the \ac{rf} carrier period. Therefore, to be able to adequately correct for the carrier phase, a further time estimation refinement process must be employed.  Typically, the matched filter is employed in the frequency domain as $s_\mathrm{mf}(t)=\mathcal{F}^{-1}\{\mathcal{F}[s_\mathrm{rx}(t)]\mathcal{F}[s_\mathrm{tx}^*(t)]\}$ to achieve a complexity of $\mathcal{O}(N\log N)$ as opposed to the $\mathcal{O}(N^2)$ required for direct convolution for an $N$-sample waveform; in this case, a zero-padding factor of $M$ may be performed on the frequency domain representation prior to the \ac{ift} which acts as an unbiased interpolation, but comes at the cost of increasing the computational complexity to $\mathcal{O}(MN\log (MN))$. To achieve high levels of interpolation using this technique, the computational load becomes intractable for real-time operation.  Thus, an alternative method must be used. Many other highly performant options exist which attempt to estimate the true time delay by using a regression model to fit a curve to the sampled matched filter data such as \ac{qls}~\cite{moddemeijer1991sampled}~\cite[\S7.2]{richards2014fundamentals}, \ac{sinc-ls}~\cite{prager2020wireless}, and \ac{mf-ls}~\cite{mghabhab2022microwave}. While \ac{mf-ls} theoretically provides an unbiased peak estimate by performing regression of the matched filter magnitude itself, it requires iterative computation of the matched filter which is typically computationally expensive, making it challenging to implement for latency-sensitive operations such as time and frequency correction. The \ac{sinc-ls} operation provides an appropriate curve fit for matched filters with a sinc-like shape such as \ac{lfm} waveforms, but is generally an iterative process, though it is noted in \cite{prager2020wireless} that a closed form solution is available for three sample points. The \ac{qls} \ac{toa} correction estimation is simple and also has a known closed form solution using three sample points~\cite[\S7.2]{richards2014fundamentals}. For a signal whose true time delay is $\tau_\mathrm{d}$ and discrete-time matched filter is $s_\mathrm{mf}[i]$ where $i$ is the sample index, the \ac{qls} process begins with an initial time estimate given simply by the peak-find result of the matched filter $i_\mathrm{max} = \mathrm{argmax}_{i\in\{0\dots2L-1\}} s_\mathrm{mf}[i]$ for a signal of length $L$, yielding the initial time estimate of $\hat{\tau}=i_\mathrm{max} / \tau_\mathrm{s,adc}$. The result is then refined by computing the centroid of a parabola fit through $i_\mathrm{max}$ and its two adjacent sample points
\begin{align}
	\label{eq:qls}
	\begin{split}
		\hat{\tau}' = \frac{\tau_\mathrm{s,adc}}{2}\frac{s_\mathrm{mf}[i_\mathrm{max}-1] - s_\mathrm{mf}[i_\mathrm{max}+1]}{s_\mathrm{mf}[i_\mathrm{max}-1]-2s_\mathrm{mf}[i_\mathrm{max}]+s_\mathrm{mf}[i_\mathrm{max}+1]} 
	\end{split}.
\end{align}%
\begin{figure}
	\centering
	\includegraphics[width=\columnwidth]{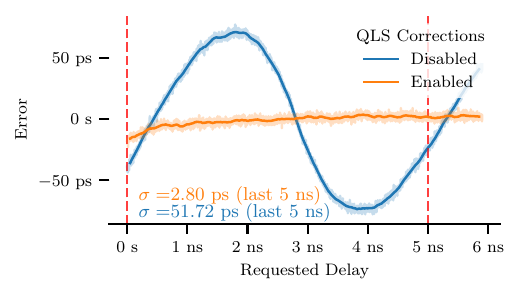}
	\caption{Time delay estimation error with and without \ac{qls} bias correction \ac{lut} applied. In this scenario a \ac{ptt} waveform with $\beta_\mathrm{ptt}=\;$\SI{40}{\mega\hertz} was transmitted from one channel and received at another on an Ettus X310 \ac{sdr} and sampled at 200\,MSa/s while the transmitted waveform was digitally delayed. Red dashed lines indicate one sample period on the \ac{adc}. The slight upward trend at the start is due to device warm-up after starting; this trend disappears after several minutes of operation.}
	\label{fig:qls-bias}
\end{figure}%
The \ac{qls} refinement value is then added to the original estimate to refine its value $\hat{\tau}_\mathrm{qls} = \hat{\tau} + \hat{\tau}'$.
Because the shape of the quadratic does not exactly match that of the matched filter, some small amount of bias is introduced in the estimate, shown experimentally in Fig. \ref{fig:qls-bias}; however, it is observed that this bias trend is a function only of the \ac{adc} sample rate and waveform parameters. Because of the cooperative nature of the system, these parameters are both known a priori allowing a \ac{lut} $\tau''(\tau')$ to be computed and used to correct for the residual \ac{qls} error as $\hat{\tau}_\mathrm{qls,lut}=\hat{\tau} + \hat{\tau}' + \hat{\tau}''$. While computing the lookup table can be slow for very fine time steps, indexing into the \ac{lut} can be performed very quickly at runtime, and a simple linear interpolation between \ac{lut} can be applied to minimize residual error~\cite{merlo2023tmtt}.

\subsection{Frequency Estimation}
\label{sec:frequency-estimation}

\begin{figure}
	\centering
	\includegraphics[width=\columnwidth]{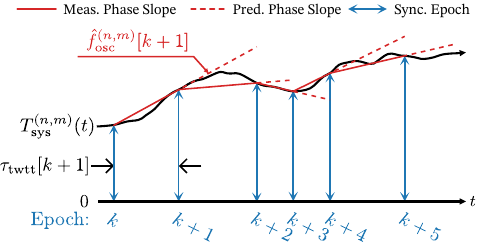}
	\caption{Frequency offset estimation process. The time transfer processes may be either uniformly sampled or arbitrarily sampled as is shown in this figure. Clock noise is shown magnified to highlight the process; in practice, the spacing between \ac{twtt} processes should be small enough that the clock drift is approximately linear.}
	\label{fig:frequency-offset-estimation}
\end{figure}


The relative frequency error can be estimated to first order as the relative phase drift divided by observation interval measured at some node $n$. As established in \eqref{eq:time-ref}, time and phase are related by the nominal frequency; thus, this can be accomplished directly by estimating the clock time offsets by using the high accuracy time estimation described in Section \ref{sec:time-estimation}, shown in Fig. \ref{fig:frequency-offset-estimation}. From this, the dimensionless frequency difference between platforms can be estimated by
\begin{align}
\begin{split}
	\Delta\hat{f}_\mathrm{osc}^{(n,m)}[k] =\frac{\hat{T}_\mathrm{sys}^{(n,m)}[k]-\hat{T}_\mathrm{sys}^{(n,m)}[k-1]}{\tau_\mathrm{twtt}[k]}
\end{split}
\end{align}
where $\tau_\mathrm{twtt}[k] = t_\mathrm{tx}^{(n)}[k]-t_\mathrm{tx}^{(n)}[k-1]$. \hlb{
Expanding using \eqref{eq:twtt-full} and \eqref{eq:internode-time-error} and letting $\delta\tau_\mathrm{tof}^{(n,m)} = \tau_\mathrm{tof}^{(n,m)}[k]-\tau_\mathrm{tof}^{(n,m)}[k-1]$ and assuming the frequency drift is quasi-static over the observation interval, i.e., $\Delta f_\mathrm{osc}^{(n,m)}[k] = \Delta f_\mathrm{osc}^{(n,m)}[k-1]$ yields
\begin{align*}
\begin{split}
\label{eq:freq-error}
	\Delta\hat{f}_\mathrm{osc}^{(n,m)}[k] 
	&= \Delta f_\mathrm{osc}^{(n,m)}[k] \left( 1 + \Delta f_\mathrm{osc}^{(n,m)}[k] \frac{\delta\tau_\mathrm{tof}^{(n,m)}}{\tau_\mathrm{twtt}[k]} \right).
\end{split}
\end{align*}
Assuming a \ac{los} path with minimal multipath, the error term can be thought of as a radial velocity and rearranged in terms of a fraction of the radial velocity over the speed of light by
\begin{align*}
	\frac{\delta\tau_\mathrm{tof}^{(n,m)}}{\tau_\mathrm{twtt}}\frac{c}{c}=\frac{\delta\tau_\mathrm{R}^{(n,m)}}{\tau_\mathrm{twtt}}\frac{1}{c}=\frac{v_R}{c}
\end{align*}
which yields
\begin{align}
	\begin{split}
		\Delta\hat{f}_\mathrm{osc}^{(n,m)}[k] &= \Delta f_\mathrm{osc}^{(n,m)}[k]\left(1+\frac{v_R}{c}\right).
	\end{split}
\end{align}
While $\frac{v_R}{c} \ll 1$, i.e., the platforms are not moving at a significant portion of the speed of light, the impact of this error will be small.
}

In a similar manner, the relative radial velocity could be estimated by the internode position change versus time $\hat{v}_r = \mathrm{d}R^{(n,m)}/\mathrm{dt}$, assuming a \ac{los} path.
Using this technique, the \ac{crlb} on frequency estimation can be approximated by \eqref{eq:var-dop} with $\zeta_t=\pi\tau_\mathrm{twtt}[k]$ where $\tau_\mathrm{twtt}[k]$ is the true time duration between the start of each synchronization epoch, assuming $\tau_\mathrm{twtt}[k] \gg \tau_\mathrm{pd}$; however, in practice, a more accurate estimate of performance can be found in a manner similar to \eqref{eq:crlb-twtt}
\begin{align}
	\begin{multlined}[\columnwidth]
		\mathrm{var}\!\left({ \Delta\hat{f}}_\mathrm{osc}^{(n,m)}[k] - \Delta f_\mathrm{osc}^{(n,m)}[k] \right) \\
		\ge \frac{1}{2\tau_\mathrm{twtt}[k]} \bigg[ \mathrm{var}\!\left( \hat{T}_\mathrm{}^{(n,m)}[k-1]- {T}_\mathrm{}^{(n,m)}[k-1]\right) \\
		+ \mathrm{var}\!\left( \hat{T}_\mathrm{}^{(n,m)}[k]- {T}_\mathrm{}^{(n,m)}[k]\right) \bigg]
	\end{multlined}
\end{align}
where the frequency estimation variance is inversely proportional to the time between observations and the average \ac{snr} of all four received waveforms, assuming the waveform parameters are held constant.


Because this technique is periodic, its accuracy will be limited by the power spectral density profile of the apparent Doppler shift, i.e., the power spectral density of platform vibrations and phase noise of the system, which have been studied in\cite{chatterjee2017a-study-of-cohe} and \cite{mghabghab2021impact-of-vco-a} for periodically synchronized systems. Critically, the update rate must be significantly greater than the frequencies in which most of the power spectrum area exists to achieve a high level of coordination.


\section{Direct Phase Compensation}
\label{sec:phase-compensation}

The uncompensated waveforms transmitted and received at node $N_n$ are given by \eqref{eq:s-tx} and \eqref{eq:s-adc}. To compensate these waveforms, the phase error imparted by the initial time offset of the system clock $T_\mathrm{0,sys}^\mathrm{(n)}$, transmitter phase delay $\phi_\mathrm{0,tx}^{(n)}$, and reference oscillator frequency offset $\delta f_\mathrm{osc}^{(n)}(t)$ which imparts both a \ac{cfo} and \ac{sfo}, need to be corrected.  This can be accomplished in two steps:
\begin{enumerate}
	\item modify the sampling time of the baseband waveform to correct the relative sampling time offset 
	\begin{align}
		\tilde{\Psi}(t_{s}[i]) = h\!\left(\Psi\!\left(t_s[i]\right); t_s[i]+ T_\mathrm{sys}^{(n,m)}(t_s[i])\right)
	\end{align}%
	where $h(x; t)$ is an arbitrary time resampling function and $\Psi(t)$ is the baseband transmit or receive waveform, and
	\item modulate the baseband waveform with a phase $\tilde{\phi}_\mathrm{lo}(t)$ opposite that of the time-varying relative internode carrier phase offsets.
\end{enumerate}
Using the linear frequency drift model, the carrier phase compensation applied at node $N_n$ is the opposite of \eqref{eq:lo-error}
\begin{align}
	\begin{split}
	\tilde{\phi}_\mathrm{lo}^{(n)}(t) &= -\phi_\mathrm{lo}^{(n,m)}(t_s[i])\\
	&= -2\pi f_\mathrm{0,lo}^{(n)} \Delta f_\mathrm{osc}^{(n,m)} t_s[i] - \phi_\mathrm{0,lo}^{(n,m)}.
	\end{split}
\end{align}
Because the analytic representation of the waveform $\Psi_\mathrm{tx}(t)$ is available at the transmitter, the underlying waveform can be directly sampled at arbitrary times, making no assumptions on the system dynamics model used; thus, the compensated transmit waveform at baseband at node $N_n$ is
\begin{align}
\label{eq:tx-correction}
\begin{split}
	\begin{multlined}[\columnwidth]
		\tilde{\Psi}_\mathrm{tx}^{(n)}[i] = \Psi_\mathrm{tx}\!\left(t_s[i]+ T_\mathrm{sys}^{(n,m)}(t_s[i])\right)\\
		\exp{-j\left[2\pi f_\mathrm{0,lo}^{(m)} \Delta f_\mathrm{osc}^{(n,m)} t_s[i] + \phi_\mathrm{0,tx}^{(n,m)}\right]}
	\end{multlined}
\end{split}
\end{align}
which, once transmitted, will cancel the time and phase offsets of the transmitter.
The compensation on the received waveform is more challenging due to the fact that there is no analytic representation available of $s_\mathrm{adc}(t)$ and thus the sampled waveform must be arbitrarily resampled to correct for \ac{sfo}. This can be modeled by
\begin{align}
\label{eq:rx-correction}
\begin{multlined}[\columnwidth]
	\tilde{\Psi}_\mathrm{rx}^{(m)}[i] = h\!\left(\Psi_{\mathrm{rx}}\!\left(t_s[i]\right); t_s[i]+ T_\mathrm{sys}^{(n,m)}(t_s[i])\right)\\
	\exp{j\left[2\pi f_\mathrm{0,lo}^{(m)} \Delta f_\mathrm{osc}^{(n,m)} t_s[i] + \phi_\mathrm{0,lo}^{(n,m)}\right]}
\end{multlined}
\end{align}
noting that the carrier phase modulation is the conjugate of that applied on the transmit side. Further challenging the receive-side sampling time compensation is the fact that truly arbitrary time resampling techniques, such as Lanczos resampling~\cite{duchon1979lanczos-filteri}, are generally computationally expensive compared to simple constant time fractional delays filters, or constant frequency offset multirate filters~\cite{harris2021multirate-signa}. However, the complexity of the receive-side sampling time compensation required is dictated by the \ac{tbp} of the incoming waveform. For long, wideband incoming waveforms, accurate compensation is necessary to ensure coherence, while for short duration and or narrowband pulses, only a constant fractional time delay may be required. This can be formulated in terms of a maximum tolerable phase error at the end of a pulse duration. In the case of the linear frequency drift model, a constant phase delay error model can be described as
\begin{align}
	\label{eq:sfo-error}
	\epsilon_\mathrm{\phi} = \Delta f \cdot \mathrm{TBP}
\end{align}
where $\Delta f$ is the dimensionless relative frequency error and TBP is the product of the incoming waveform duration and its bandwidth (i.e., the time-bandwidth product).  The maximum tolerable $\epsilon_\mathrm{\phi}$ will depend heavily on the application. As an example, in the case of \ac{twtt}, the frequency error of the incoming waveform will impact the matched filter response, as shown by the ambiguity function in Fig. \ref{fig:ambiguity-function}, given by \eqref{eq:ambiguity-function}, if the total frequency error is less than $\beta_\mathrm{pd}/2$, the peak find will still obtain the correct time result in a noise-free scenario; in this case, the incoming waveform may not need to be resampled, instead only the final \ac{toa} estimate may be compensated, saving significant computation time. In this work, the system has typical internode frequency errors of $\Delta f <$ 1\,ppm and the waveform used has a \hbox{\ac{tbp} $ = 30$} yielding $\epsilon_\phi=$\,\SI{1.8}{\milli\degree} of accumulated phase error, thus having little impact on the performance.


\section{Wireless Coordination Experiments}
\label{sec:wireless-coordination-experiment}

\subsection{Hardware Configuration}

The wireless coordination experiments consisted of three different hardware configurations. In all three configurations, each node consisted of an Ettus Research X310 \acf{usrp} connected to a Dell Optiplex 7080 host with an Intel i7-10700 and 16\,GB of DDR4 memory connected via 10\;Gigabit Ethernet (GbE). The hosts were connected via 1\;GbE to a central network switch for digital communication between nodes. The radios on each node utilized channel A to perform \ac{twtt} for time and frequency offset estimation; the first and last nodes in the array (node $N_0$ and $N_1$) contained an external Mini-Circuits ZFSWA2R-63DR+ antenna diversity switch to multiplex between a \ac{twtt} antenna, and a ``beamforming'' cable connected to an oscilloscope to directly measure beamforming signals from each node at an independent external observer to characterize expected real-world beamforming performance parameters and isolate time, frequency, and phase performance of each node. The \ac{twtt} antenna used in all experiments was a Taoglas TG.56.8113 wideband monopole. In all configurations, the antenna on node $N_0$ was affixed to a motorized linear actuator. In static configurations, the antenna $N_0$ was held fixed at a distance of $\sim$\SI{1}{\meter} from antenna $N_1$; in dynamic experiments the internode distance between $N_0$ and $N_1$ varied from \SI{37}{\centi\meter} to \SI{134}{\centi\meter}, spanning $\sim$\SI{1}{\meter}. The three hardware configurations were designated \textit{Configuration A} through \textit{Configuration C}.
\begin{figure*}
	\centering
	\includegraphics[width=0.31\textwidth]{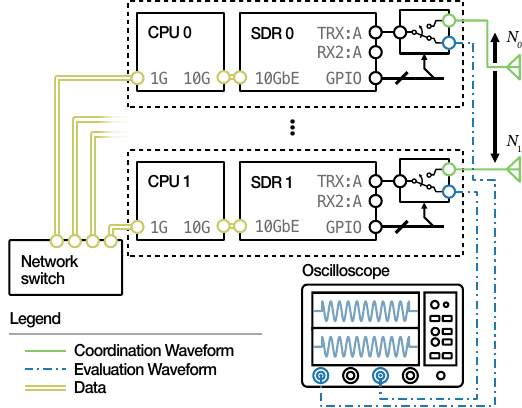}\hfill{}
	\includegraphics[width=0.31\textwidth]{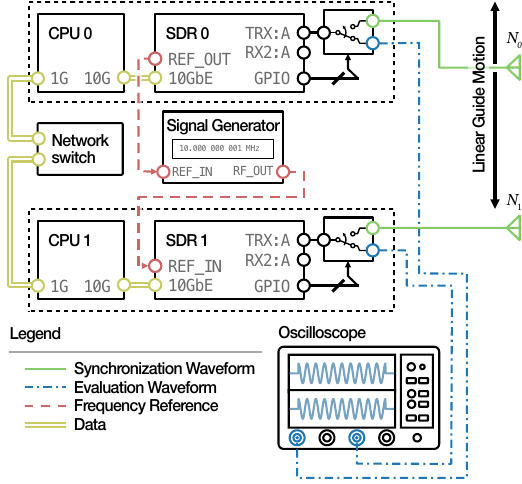}\hfill{}
	\includegraphics[width=0.31\textwidth]{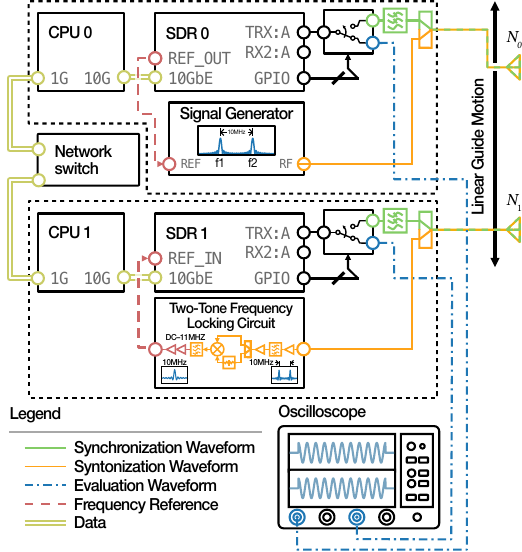}\\
	(a)\hspace{0.33\textwidth}(b)\hspace{0.33\textwidth}(c)
	\caption{(a) \textit{Configuration A:} Fully-digital wireless time and frequency coordination.  Number of nodes was varied between 2 and 4 in various experiments. In all experiments, nodes $N_0$ and $N_1$ were connected to the oscilloscope; due to the largest \ac{tdma} time separation and largest internode distance, this evaluates the worst-case synchronization performance in the array. (b) \textit{Configuration B:} Fully-digital time and frequency coordination with a known static relative frequency offset via signal generator.  In this experiment, the frequency offset estimation accuracy is evaluated using the known frequency error introduced by the signal generator. (c) \textit{Configuration C:} Fully-digital time coordination with analog continuous wireless frequency transfer based on~\cite{merlo2023tmtt}. }
	\label{fig:configuration}
\end{figure*}
\begin{itemize}
	\item \textit{Configuration A} (Fig. \ref{fig:configuration}a) was configured to run using the fully wireless digital time-frequency coordination procedure described in this paper with varying parameters.  Parameters evaluated were \ac{snr}, pulse duration $\tau_\mathrm{pd}$, epoch duration $\tau_e$, average resynchronization interval $\tau_\mathrm{twtt}$, and node scaling from 2--4 nodes under static and dynamic cases. 
	\item \textit{Configuration B} (Fig. \ref{fig:configuration}b) evaluated the ability of the system to estimate and correct for a known fixed frequency offset. This configuration added a fixed frequency offset between nodes $N_0$ and $N_1$ by using the \SI{10}{\mega\hertz} frequency reference from the \ac{sdr} of node $N_0$ to discipline a Keysight N5183A MXG signal generator signal generator which was used to generate a reference signal with a known frequency error which was provided to the \SI{10}{\mega\hertz} frequency reference input on the \ac{sdr} of node $N_1$. Because the oscillator noise is no longer independent between nodes, this allows a more accurate lower bound of the estimator accuracy for the system to be evaluated under a known frequency offset, independent of system oscillator drifts. 
	\item \textit{Configuration C} (Fig. \ref{fig:configuration}c) evaluated the performance of the system compared with the two-tone \ac{cw} analog wireless frequency transfer technique used in \cite{merlo2023tmtt,abari2015airshare,mghabghab2021open-loop-distr} under relative motion with varying velocities. A Keysight E8267D PSG signal generator disciplined by the reference oscillator on \ac{sdr} 0 was used to transmit the \ac{cwtt} frequency reference. In this configuration, the frequency transfer is handled entirely via the analog system, and thus the \ac{sdr} digital \ac{rf} system is only used for time and range estimation. Because the frequency transfer circuit directly disciplines the SYS PLL in this scenario, the phase will also change based on the \ac{tof} between the transmitting and receiving nodes. To correct for this phase rotation, the \ac{twtt} estimate is used to apply an additional compensation phase in this configuration.
\end{itemize}
\begin{figure}
	\centering
	\includegraphics[width=\columnwidth]{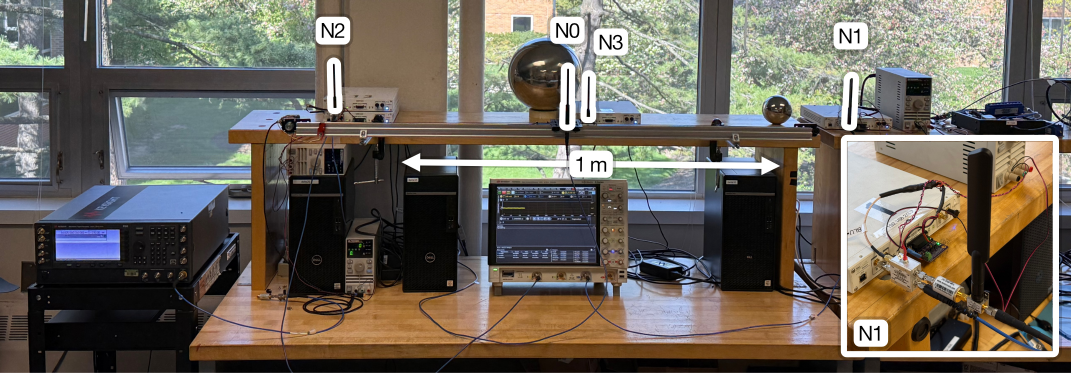}
	\caption{Photograph of the experimental configuration.  Nodes were numbered $N_0$ through $N_3$; nodes $N_0$ and $N_1$ were always connected to the oscilloscope to measure beamforming performance.}
	\label{fig:experimental-setup-photo}
\end{figure}
In \textit{Configuration A}, a Keysight DSOS8404A oscilloscope was used, and in \textit{Configuration B} and \textit{Configuration C}, a Keysight MSOX92004A oscilloscope was used. A photograph of the experimental setup is show in Fig. \ref{fig:experimental-setup-photo}. \hlb{To study the impact of multipath on the coordination, two metallic spheres were also included near the antennas to induce strong time-varying multipath scattering as the antenna was moved.}

\subsection{Software Configuration}


The software was developed using \ac{usrp} Hardware Driver (UHD) 4.8 and GNU Radio 3.10 and ran on Ubuntu 24.04. The software was separated logically into three separate programs termed \textit{controllers} shown schematically in Fig. \ref{fig:software-diagram}. In this system, each  controller was a GNU Radio flow graph. The blocks in each flow graph were primarily written using Python 3.12, making extensive use of NumPy 
and Numba 
to achieve high performance, with only data manipulation blocks written in C++. 
\begin{itemize}
	\item\textit{Node Controller:} This controller acts as the primary arbitrator between other array controllers and the \ac{usrp} resource. It interprets waveform description dictionaries and samples the waveform on-device in an \acf{awg} block, using the known sampling time offsets and applies a beat frequency to compensate for the \ac{lo} frequency error, as described by \eqref{eq:tx-correction}. On the receive side it performs \ac{cfo}, \ac{sfo}, and phase corrections, as needed, described by \eqref{eq:rx-correction}; in these experiments the \ac{sfo} was not enabled on receive due to the minimal amount of error in the \ac{twtt} messages, as described by \eqref{eq:sfo-error} in Section \ref{sec:phase-compensation}. Finally, this controller also handles the \ac{toa} estimation by performing the matched filter and \ac{qls}--\ac{lut} refinement process required for high-accuracy \ac{twtt}, described in Section \ref{sec:high-accuracy-toa}, distributing the processing across nodes and performing data compression so only the timestamps must be transmitted between nodes.
	
	\item\textit{Synchronization Controller:} This controller performs the \ac{twtt} process described in Section \ref{sec:twtt} by orchestrating the Node Controllers to transmit \ac{twtt} waveforms at known times and collecting their refined \acp{toa} estimates; these estimates are then used to compute the frequency offsets using the techniques described in Section \ref{sec:frequency-estimation}.
	
	\item\textit{Beamforming Controller:} This controller generates the evaluation pulses that are sent to the oscilloscope to verify system performance.   	
	
\end{itemize}
Within each GNU Radio flow graph, interprocess communication was accomplished using the message passing interface instead of the streaming interface to reduce latency at the expense of overall computational bandwidth; similarly, the radios were operated via UHD in a ``bursty'' mode, using finite duration pre-scheduled timed transmit and receive commands to ensure the exact transmit and receive times are known including host to radio transport latency. Between each of the controllers, communication was accomplished using ZeroMQ via TCP/IP to scale the distributed computing model easily between hosts. In this system the Beamforming and Synchronization controllers are centralized and are run on host 0 in a \textit{centralized} manner; however, it should be noted that \textit{decentralized} time and frequency transfer could be implemented using average consensus-based algorithms~\cite{ouassal2020decentralized,shandi2024decentralized-p}. 


This work also uses an iterative time refinement estimation technique on device startup which uses network timestamps from each host to obtain a coarse level of refinement ($\sim$\SI{10}{\milli\second}). 
Initially, the time on each radio in the array is queried and returned over the network via TCP/IP which has an uncertainty on the order of $\sim$\SI{10}{\milli\second}. This determines the initial \ac{tdma} window duration $\tau_\mathrm{tdma}$. Because the window is quite large at initialization, the sample rate begins at a low rate, and thus the \ac{ptt} tone separation $\beta_\mathrm{ptt}$ is also kept low. After the initial network time alignment, the synchronization index \texttt{sync\_idx} is incremented with each \ac{twtt}, which controls the sample rate, tone separation, and \ac{tdma} window size. Table \ref{tab:time-refinement-parameters} summarizes the values used for these parameters in this work. This eliminates the need for more accurate forms of synchronization such as \acf{pps}, typically derived from \acf{gnss} sources which may not always be available; because this technique uses TCP/IP, it can also be incorporated over Wi-Fi or other wireless IP-based links, demonstrated previously~\cite{merlo2024past}.

\subsection{Performance Evaluation Methods}
\label{sec:methods}

To evaluate the performance of the system in each of the experiments, nodes $N_0$ and $N_1$ were connected to an oscilloscope and ``beamforming'' pulses were transmitted coherently via coaxial cables to evaluate the performance at an external location. These beamforming pulses were sampled at the oscilloscope at 20\,GSa/s and the interarrival time, phase, and frequency were estimated and the signals were digitally summed together as they would be when beamforming over the air. In the experiments, four quantities were derived from the beamforming measurements:
\begin{enumerate}
	\item \textit{Coherent Gain:} a measure of coherence, computed by comparing the power of the summed received waveform relative to the sum of the power of the individual waveforms given by
\begin{align}
	\label{eq:coherent-gain}
	G_c = \frac{\sum_{i=0}^{L}\left|s_\mathrm{rx}^{(0)}[i]+s_\mathrm{rx}^{(1)}[i]\right|^2}
	{2\left(\sum_{i=0}^{L}\left|s_\mathrm{rx}^{(0)}[i]\right|^2+\left|s_\mathrm{rx}^{(1)}[i]\right|^2\right)}
\end{align}
where $s_\mathrm{rx}^{(0)}$ and $s_\mathrm{rx}^{(1)}$ are the waveforms received from nodes 0 and 1, respectively and $L$ is the number of samples of each waveform.
\item \textit{Interarrival Time:} the time between each waveform arriving at the channels of the oscilloscope, computed by interpolating the cross-correlation $s_\mathrm{xy}(t)=|\mathcal{F}^{-1}\{\mathcal{F}[s_\mathrm{rx}^{(0)}(t)]\mathcal{F}[s_\mathrm{rx}^{*(1)}(t)]\}|$ peak output using \ac{qls} as described in \eqref{eq:qls}.
\item \textit{Interarrival Phase:} the phase difference between the waveforms arriving at the oscilloscope, computed by estimating the relative phase difference at each sample whose magnitude was above $0.9\cdot \max{|s_\mathrm{rx}^{(n)}|}$.
\item \textit{Internode Frequency Difference:} the frequency difference between waveforms arriving at the oscilloscope, computed using the weighted relative phase averaging technique described in~\cite{kay1989a-fast-and-accu} where the phase difference between each sample measured at the oscilloscope over a \SI{100}{\micro\second} \ac{cw} pulse, trimming \SI{1}{\micro\second} of data from the start and end of the waveform to avoid transient signal distortion due to rising and falling edges.
\end{enumerate}
In addition, to the measured beamforming internode frequency difference measurements, the estimated frequency offset $\Delta\hat{f}_\mathrm{osc}$ computed at Node $N_1$ is also presented for comparison. In all experiments, the \ac{snr} of the waveform was estimated by comparing the energy of the signal where the waveform was present, to the energy immediately after the waveform envelope
\begin{align}
	\mathrm{SNR}=10\log_{10}\!\left[\frac{E_s}{E_n}\right]=10\log_{10}\!\left[\frac{\mathrm{max}_{i\in\{0\dots 2L-1\}} s_\mathrm{mf}[i]}{\frac{1}{L_n}\sum_{i=0}^{L_n}|\nu_\mathrm{rx}[i]|^2}\right]
\end{align}
where $E_s$ is the signal power of the $L$ received waveform samples, and $E_n$ is the signal power of the $L_n$ noise samples.

\subsection{Experimental Results}
\label{sec:experimental-results}

\begin{table}
	\caption{Waveform Parameters}
	\label{tab:experiment-parameters}
	\begin{center}
  	\begin{tabularx}{\columnwidth}{p{0.25\linewidth}p{0.1\linewidth}YYY}
	\toprule[1pt]
	\multicolumn{2}{l}{\textbf{Nominal Time Transfer Waveform}} \\
	\midrule
	\midrule
	\multirow{2}{*}[-0.4em]{\bf{Parameter}} & \multirow{2}{*}[-0.4em]{\bf{Symbol}} & \multicolumn{3}{c}{\bf{Configuration}}\\
	\cmidrule{3-5}
	&  & A & B & C \\
	\midrule
	Waveform Type &  & \multicolumn{3}{c}{\centeredline{\mbox{\acs{ptt}}}} \\
	Carrier Frequency & $f_{0,\mathrm{ptt}}$ & \SI{2.1}{\giga\hertz} & \SI{2.1}{\giga\hertz} & \SI{700}{\mega\hertz} \\
	Max. Tone Separation & $\beta_\mathrm{ptt}$ & \multicolumn{3}{c}{\centeredline{\SI{20}{\mega\hertz}}} \\
	Rise/Fall Time & & \multicolumn{3}{c}{\centeredline{\SI{50}{\nano\second}}} \\
	Pulse Duration & $\tau_\mathrm{pd}$ & \multicolumn{3}{c}{\centeredline{\SI{1.5}{\micro\second}}} \\
	Synchronization Epoch Duration & $\tau_\mathrm{e}$ & \multicolumn{3}{c}{\centeredline{\SI{11.5}{\micro\second}}} \\
	Resynchronization Interval & $\tau_\mathrm{twtt}$ & \multicolumn{3}{c}{\centeredline{$\sim$\SI{40}{\milli\second}}} \\
	Tx Sample Rate & $f_\mathrm{s,dac}$ & \multicolumn{3}{c}{\centeredline{\SI{200}{\mega Sa/s}$^*$}} \\
	Rx Sample Rate & $f_\mathrm{s,adc}$ & \multicolumn{3}{c}{\centeredline{\SI{200}{\mega Sa/s}}}\\
	\midrule[1pt]
	\multicolumn{2}{l}{\textbf{Frequency Transfer Waveform}}\\
	\midrule
	\midrule
	Parameter & Symbol & A & B & C \\
	\midrule
	Waveform Type &  & --- & --- & \acs{cw}TT \\
	Carrier Frequency & $f_{0,\mathrm{f}}$ & --- & --- & \SI{2.1}{\giga\hertz} \\
	Tone Separation & $\beta_\mathrm{f}$ & --- & --- & \SI{10}{\mega\hertz} \\
	\midrule[1pt]
	\multicolumn{5}{l}{\textbf{Beamforming Waveform (Time \& Phase Measurement)}} \\
	\midrule
	\midrule
	Parameter & Symbol & A & B & C \\
	\midrule
	Waveform Type &  & \multicolumn{3}{c}{\centeredline{\ac{ptt}}} \\
	Carrier Frequency & $f_{0,\mathrm{bf}}$ & \multicolumn{3}{c}{\centeredline{\SI{1.0}{\giga\hertz}}} \\
	Rise/Fall Time & & \multicolumn{3}{c}{\centeredline{\SI{50}{\nano\second}}} \\
	Pulse Duration & $\tau_\mathrm{pd}$ & \SI{2.0}{\micro\second} & \SI{1.0}{\micro\second} & \SI{1.0}{\micro\second} \\
	Tx Sample Rate & $f_\mathrm{s,dac}$ & \multicolumn{3}{c}{\centeredline{\SI{200}{\mega Sa/s}$^*$}} \\ 
	Rx Sample Rate & $f_\mathrm{s,osc}$ & \multicolumn{3}{c}{\centeredline{\SI{20}{\giga Sa/s}}} \\
	\midrule[1pt]
	\multicolumn{5}{l}{\textbf{Beamforming Waveform (Frequency Measurement)}} \\
	\midrule
	\midrule
	Parameter & Symbol & A & B & C \\
	\midrule
	Waveform Type &  & \multicolumn{2}{c}{\centeredline{\ac{cw}}} & --- \\
	Carrier Frequency & $f_{0,\mathrm{bf}}$ & \multicolumn{2}{c}{\centeredline{\SI{1.0}{\giga\hertz}}} & --- \\
	Rise/Fall Time & & \multicolumn{2}{c}{\centeredline{\SI{50}{\nano\second}}} & --- \\
	Pulse Duration & $\tau_\mathrm{pd}$ & \multicolumn{2}{c}{\centeredline{\SI{100}{\micro\second}}} & ---  \\
	Tx Sample Rate & $f_\mathrm{s,dac}$ & \multicolumn{2}{c}{\centeredline{\SI{200}{\mega Sa/s}$^*$}} & --- \\ 
	Rx Sample Rate & $f_\mathrm{s,osc}$ & \multicolumn{2}{c}{\centeredline{\SI{20}{\giga Sa/s}}} & ---\\
	\bottomrule[1pt]
	\end{tabularx}
	\end{center}
	$^*$ Digitally interpolated from \SI{200}{\mega Sa/s} to \SI{800}{\mega Sa/s} on the \ac{dac}
\end{table}

Several experiments were conducted to assess the performance of the system under varying waveform and system parameters.  The nominal waveform parameters used were varied based on system configuration and are summarized in Table \ref{tab:experiment-parameters}. Notably, a \SI{50}{\mega\hertz} \ac{ptt} waveform was used for beamforming evaluation in the coherent gain, interarrival time, and interarrival phase measurements due to its superior ability to estimate interarrival time and phase, described in Section \ref{sec:high-accuracy-toa}; a \SI{100}{\micro\second} pulsed \ac{cw} waveform was used in the internode frequency estimation measurements for improved frequency estimation ability. Prior to conducting the experiments, a phase calibration was performed to address phase shifts due to small mismatches in transmission line lengths between and phase delays in the \acp{rfe} of the \acp{sdr} and the oscilloscope. The same phase calibration was used in \textit{Configuration A} and \textit{Configuration B}; however, recalibrations occurred prior to conducting experiments using \textit{Configuration C} due to changes in the \ac{rfe} components, and prior to frequency estimation due to the measurements occurring on a different day. No time delay calibrations were performed in these experiments which resulted in a static delay error of $\sim$\SI{100}{\pico\second} that could be further calibrated to improve system alignment performance; the impact of this small delay is minor due to the envelope modulation of only \SI{50}{\mega\hertz} due to the \ac{ptt} waveform used for beamforming, resulting in $\sim$0.5\,\% envelope time error. In all experiments, the system would alternately perform a single \ac{twtt} exchange between all nodes, then perform a beamforming pulse, in an infinite loop. In the internode frequency difference experiments, only a small subset of the beamforming pulses were recorded due to the long recording times on the scope.

To summarize the statistics in the experiments measuring the coherent gain, interarrival time, and interarrival phase, 1014 pulses of data were collected over $\sim$\SI{50}{\second} and letter-value plots are used to visualize the data. These plots are designed to illustrate large non-normally distributed datasets~\cite{hofmann2017value-plots-bo}. In these figures, the central bar represents the median of the data, and the central box represents 50\,\% of the data; each successive box thereafter represents half of the remaining data (e.g., 25\,\%, 12.5\,\%, etc.). The boxes extend to approximately 1.5 times the interquartile range, similar to the ``whiskers'' on box plots, where data beyond this value is considered an outlier and indicated by a circle marker; outlier marks are omitted in the ``time difference'' plots due to a small number of significant outliers which occur when the \ac{twtt} timing is not met by the host processor, causing the \ac{twtt} to be retried. Implementing a real-time processing scheme on the host, or moving the computation to a real-time processor, such as the \acf{fpga} on the \acp{sdr}, would mitigate this issue. In the internode frequency measurements $\sim$54 pulses were collected and standard box plots are used to represent the data; however, due to limitations on the oscilloscope in the data collection process used, the system performance could not be directly associated with each beamforming pulse saved on the oscilloscope, so the aggregated statistics (e.g., \ac{snr}) over the full run are reported instead.

In all experiments, standard deviation values are also plotted with the distributions of the measured data. Because the standard deviation is not a robust measure of statistical dispersion, outliers due to the host processor missing real-time processing deadlines causing synchronization retries are removed in the experiments with unbounded measurement values (interarrival time and internode frequency difference) prior to computing the standard deviation. Outliers are removed by iteratively removing data which exceeds $x$ standard deviations until all data fits within $x$ standard deviations; for the interarrival time dataset, $x=6$; for the smaller frequency difference dataset, $x=4$.
Additionally, due to the significantly longer pulses used in frequency estimation, the host processing required to generate and transfer all the samples to the \acp{sdr} increased appreciably, resulting in longer resynchronization intervals and decreasing overall coordination performance. Nonetheless, the frequency syntonization results still indicate good performance. Finally, statistics from each run are summarized in time domain plots and histograms, with markers indicating where outliers have been removed, in the appendices located in the supplemental materials.


\begin{figure*}
	\centering
	\includegraphics[width=0.3\textwidth]{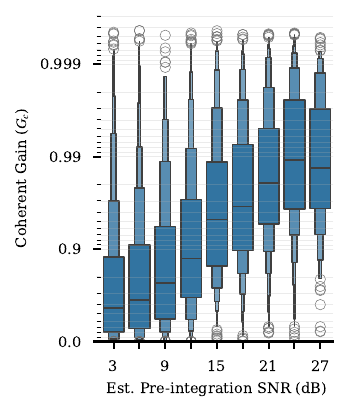}\hfill{}
	\includegraphics[width=0.3\textwidth]{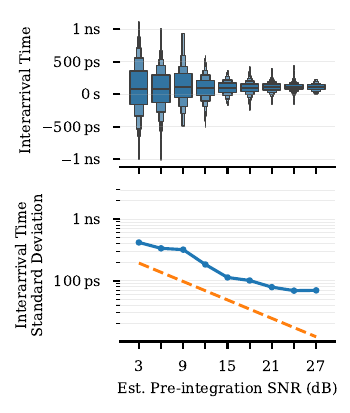}\hfill{}
	\includegraphics[width=0.3\textwidth]{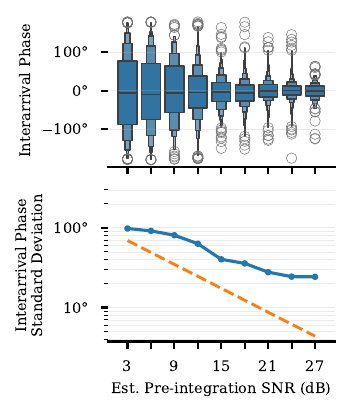}
	\caption{Electrical state coordination performance evaluation under varying average pre-integration \ac{snr} of the received waveform measured at each node during the \ac{twtt} process. The top row shows the absolute measured quantity, and the bottom row shows its standard deviation. In the interarrival time standard deviation, the outliers have been removed as discussed in Section \ref{sec:methods}. The dashed line indicates the \acf{crlb}.} 
	\label{fig:snr}
\end{figure*}

\hlb{In addition, to verify the multipath effect imparted by the metallic spheres, the \ac{tof} was computed and plotted as the linear guide traversed the \SI{1}{\meter} path relative to the unwrapped internode \ac{twtt} carrier phase difference to indicate the approximate linear guide trajectory, shown in Fig. \ref{fig:multipath}.  In this figure the uncalibrated \ac{tof} is used which includes static delay terms due to \ac{dsp} delays, filter group delays, and propagation delays through transmission lines, thus the absolute \ac{tof} cannot be provided, only a relative estimate.}

\subsubsection{Signal-to-Noise Ratio}

 \begin{figure}
	\centering
	\includegraphics[width=0.9\columnwidth]{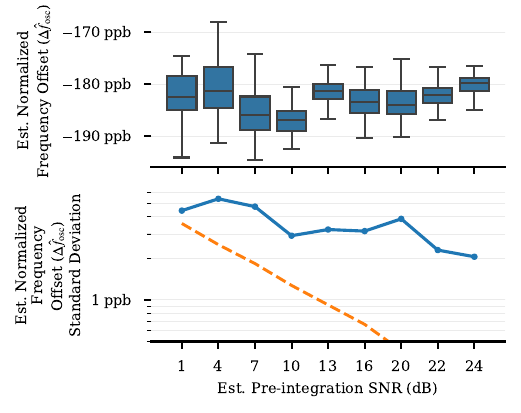}
	\caption{Estimated internode frequency offset and standard deviation under varying nominal pre-integration \ac{snr} of the received waveform measured at each node during the \ac{twtt} process. The top row shows the absolute measured quantity, and the bottom row shows its standard deviation, omitting outliers as discussed in Section \ref{sec:methods}. The \ac{crlb}, given by \eqref{eq:var-dop} using the nominal \ac{snr} and average $\tau_\mathrm{twtt}=\SI{55}{\milli\second}$, is indicated by the dashed line.}
	\label{fig:snr-est-freq}
\end{figure}
\begin{figure}
	\centering
	\includegraphics[width=0.9\columnwidth]{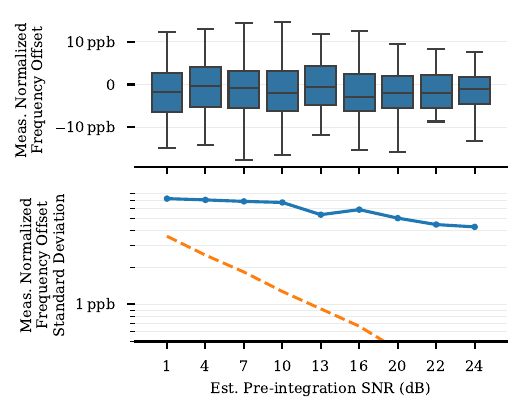}
	\caption{Measured internode ``beamforming'' frequency offset relative to node 0 at oscilloscope versus the nominal pre-integration \ac{snr} of the received \ac{ptt} waveform at each node during the \ac{twtt} process. The top row shows the absolute measured quantity, and the bottom row shows its standard deviation, omitting outliers as discussed in Section \ref{sec:methods}. The \ac{crlb}, given by \eqref{eq:var-dop} using the nominal \ac{snr} and average $\tau_\mathrm{twtt}=\SI{55}{\milli\second}$, is indicated by the dashed line.}
	\label{fig:snr-meas-freq}
\end{figure}

In this experiment \textit{Configuration A} was used to evaluate the coherent gain, interarrival time, and interarrival phase (Fig. \ref{fig:snr}), and estimated and measured internode beamforming frequency (Fig. \ref{fig:snr-est-freq} and Fig. \ref{fig:snr-meas-freq}) under varying average pre-integration \ac{snr} (i.e., before filtering) for the \ac{twtt} \ac{ptt} waveform received at each node. As the average \ac{snr} increases, the coordination performance improves, which can be seen in the median coherent gain which increases to a maximum where the \ac{snr} reaches 24\,dB, 
 at which point a limit of about 0.99 is reached due to the static time and phase calibration values.  The time difference of arrival has a relative constant median near $\sim$\SI{100}{\pico\second} due to unmatched static system delays, but the data spread decreases with \ac{snr}, as expected approaching a timing standard deviation of \SI{70}{\pico\second}. The interarrival phase maintained a zero-mean offset with a standard deviation of \hlb{$\sim$\ang{23}} at 27\,dB \ac{snr}. 

The estimated and measured frequency syntonization performance plots are shown in Figs. \ref{fig:snr-est-freq} and \ref{fig:snr-meas-freq}, respectively. The absolute frequency offset estimates are shown on the top row of Fig. \ref{fig:snr-est-freq}, which indicates a typical internode frequency offset of near $-182$\,ppb and estimated frequency offset standard deviation of 2.04\,ppb. The beamforming frequency offsets after compensation are shown in Fig. \ref{fig:snr-meas-freq} which indicate a mean frequency error of $-1.25$\,ppb across all \acp{snr} with the standard deviation reaching $4.22$\,ppb at a 30\,dB \ac{snr}. 
 The \ac{crlb} is also included using \eqref{eq:var-dop} as indicated by the dashed line. This \ac{crlb} is computed using the nominal \ac{snr} value and the average resynchronization period $\tau_\mathrm{twtt}=\SI{55}{\milli\second}$.

\subsubsection{Reference Oscillator Frequency Offset}

\begin{figure*}
	\centering
	\includegraphics[width=0.3\textwidth]{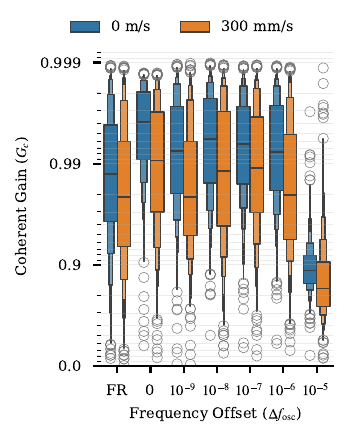}\hfill{}
	\includegraphics[width=0.3\textwidth]{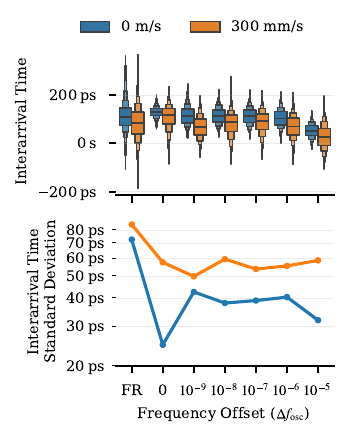}\hfill{}
	\includegraphics[width=0.3\textwidth]{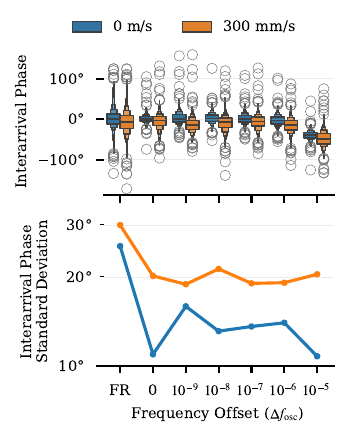}
	\caption{Electrical state coordination performance evaluation using known static frequency offsets generated using a signal generator, as described in \textit{Configuration B}. The top row shows the absolute measured quantity, and the bottom row shows its standard deviation, omitting outliers as discussed in Section \ref{sec:methods}. The data point labeled ``FR'' indicates both oscillators are ``free-running,'' i.e., not using an external frequency reference. Across all measurements, an average pre-integration \ac{snr} of approximately $\mu=24$\,dB and $\sigma=1$ was estimated.} 
	\label{fig:frequency-offset}
\end{figure*}
\begin{figure}
	\centering
	\includegraphics[width=0.9\columnwidth]{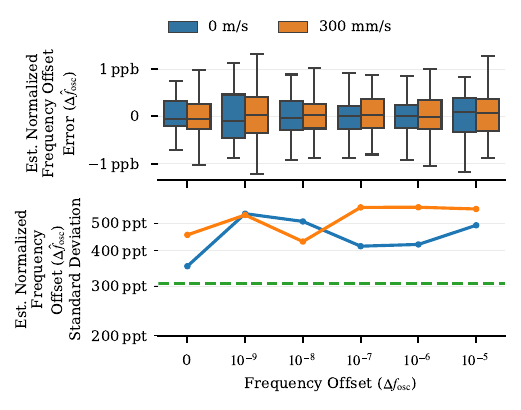}
	\caption{Estimated frequency offset between nodes (top) and its standard deviation (bottom) using \textit{Configuration B}. The top row shows the absolute measured quantity, and the bottom row shows its standard deviation, omitting outliers as discussed in Section \ref{sec:methods}. The \ac{crlb} is indicated by the dashed line. Across all measurements the \ac{snr} was estimated to have a mean value of 22.5\,dB 
	with a resynchronization interval of $\tau_\mathrm{twtt}=\SI{55}{\milli\second}$ Note that ``FR'' is omitted as this plot represents estimator error and the true value of the frequency error between nodes is unknown.}
	\label{fig:frequency-offset-est-freq} 
\end{figure}
\begin{figure}
	\centering
	\includegraphics[width=0.9\columnwidth]{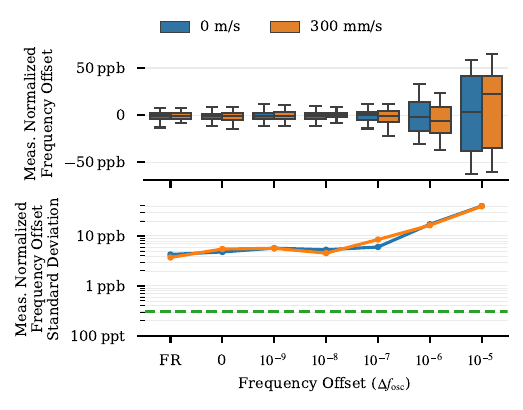}
	\caption{Internode beamforming frequency offsets measured at the oscilloscope in \textit{Configuration B}. The top row shows the absolute measured quantity, and the bottom row shows its standard deviation, omitting outliers as discussed in Section \ref{sec:methods}.  The data point labeled ``FR'' indicates both oscillators are ``free-running.''}
	\label{fig:frequency-offset-meas-freq}
\end{figure}

This experiment used \textit{Configuration B} to evaluate the accuracy of the system when estimating a known internode frequency offset $\Delta f_\mathrm{osc}^{(n,m)}$ by locking the system clocks together with a known static frequency offset generated by a signal generator. Known frequency offsets ranging from 0 to 10\,ppm and free-running (FR) were evaluated under static and dynamic internode positions and the performance results are summarized in Figs. \ref{fig:frequency-offset}--\ref{fig:frequency-offset-meas-freq}. There is a minimal change in coherent gain, time, and phase stability up to frequency offsets of 1\,ppm with performance drop-off at 10\,ppm due primarily to increased time and phase biases.

Figs. \ref{fig:frequency-offset-est-freq} and \ref{fig:frequency-offset-meas-freq} show the estimated and measured frequency offsets, respectively, along with their standard deviations, omitting outliers due to the host processor not meeting real-time requirements as discussed in Section \ref{sec:methods}, and \ac{crlb} from \eqref{eq:var-dop} using the average \ac{snr} across all measurements of 19.5\,dB 
 and average resynchronization period of $\tau_\mathrm{twtt}=\SI{55}{\milli\second}$. The results for the estimated frequency error are shown in Fig. \ref{fig:frequency-offset-est-freq} which appears as zero-mean over all frequency offsets with minimal standard deviation. This trend is continued for the beamforming frequency errors measured at the oscilloscope, shown in Fig. \ref{fig:frequency-offset-meas-freq}, up to 1\,ppm. After investigating the internode phase differences in the time domain (See experiments 26.6 and 26.13 Appendix \ref{ap:time-domain-frequency-offset} in the supplemental materials), the source of this error appears due to a harmonic sawtooth-like phase error which evolves on the order of microseconds, likely due to the \ac{pll} designs, and thus cannot be compensated for using the resynchronization period on the order of milliseconds used in these experiments.

\subsubsection{Frequency Syntonization Method}
\label{sec:exp-syntonization-method}
\begin{figure*}
	\centering
	\includegraphics[width=0.3\textwidth]{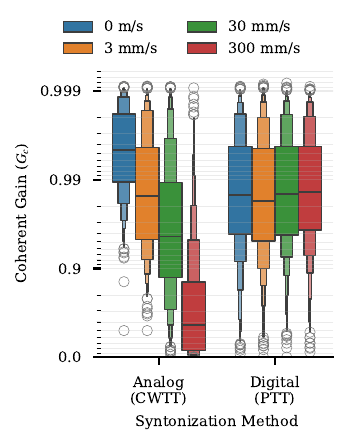}\hfill{}
	\includegraphics[width=0.3\textwidth]{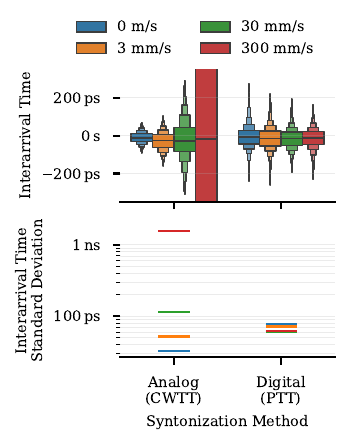}\hfill{}
	\includegraphics[width=0.3\textwidth]{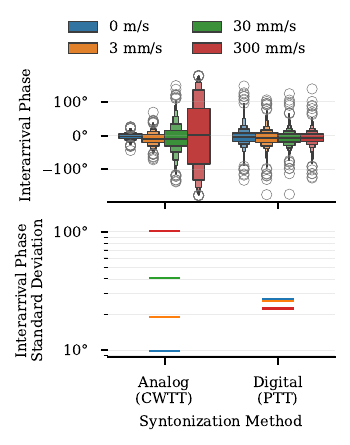}
	\caption{Electrical state coordination performance evaluation for the analog \acf{cwtt} method versus the \ac{twtt}-based fully-digital \acf{ptt} approach described in this paper under varying relative motion. The top row shows the absolute measured quantity, and the bottom row shows its standard deviation, omitting outliers as discussed in Section \ref{sec:methods}.  The clipped data for \SI{300}{\milli\meter/\second} interarrival time spans over approximately $\pm$\SI{2.5}{\nano\second}, or $\pm$0.5 samples at \SI{200}{\mega Sa/\second}.}
	\label{fig:syntonization-method}
\end{figure*}

This experiment used \textit{Configuration C} to evaluate the \ac{twtt}-based fully-digital \ac{ptt} time and frequency estimation technique described in this work relative to the analog \acf{cwtt}-based  syntonization technique used in prior works~\cite{abari2015airshare,mghabghab2021open-loop-distr,merlo2023tmtt,merlo2024ims}. In this technique the analog \ac{cwtt} method is used for syntonization and the digital \ac{twtt} method is used to compensate for the time-varying phase shift induced by the time-varying propagation delay of the \ac{cwtt} clock signal received at the secondary nodes when the system is in motion. In this experiment, the linear guide was set to \SI{0}{\meter/\second}, \SI{3}{\milli\meter/\second}, \SI{30}{\milli\meter/\second}, and \SI{300}{\milli\meter/\second}, with a range of movement of \SI{0}{\centi\meter}, \SI{1}{\centi\meter}, \SI{10}{\centi\meter}, and \SI{100}{\centi\meter}, respectively, centered about a distance of $\sim$\SI{87}{\centi\meter} between nodes 0 and 1. While both systems perform satisfactorily in stationary scenarios, the analog \ac{cwtt} frequency syntonization technique maintains the highest performance yielding nearly all measurements above $G_c>0.9$; in slowly moving scenarios of \SI{3}{\milli\meter/\second}, mimicking situations with slight oscillations in a static installment, performance is comparable between the two systems. However, in situations with more significant relative motion, the digital \ac{ptt} \ac{twtt}-based system performs significantly better. This is primarily due to the fact that phase of the received frequency reference varies rapidly due to the changing propagation delay and multipath in the environment; while the \ac{twtt} process corrects for this periodically, even at modest speeds the received phase varies significantly faster than the rate of the \ac{twtt} required to correct for unlocked oscillators resulting in lower performance overall. Furthermore, due to environmental multipath, this phase modulation on the received frequency reference waveform is difficult to predict due to platform motion alone, thus tracking this phase via kinematic models may also be challenging.

\subsection{Discussion}

Many prior works have been referenced demonstrating time and frequency coordination in open-loop wirelessly coordinated \acp{cda}; however, due to varying performance measurement techniques and reporting of coordination parameters used, a direct comparison between works is challenging. Nontheless, a tabular comparison of a selection of similar methods with similar parameter and performance reporting methodologies is included in Table \ref{tab:time-transfer-comparision}. A summary of both the hybrid and fully digital techniques are included at the bottom of the table with the fully digital \ac{ptt}-based technique is shown in boldface font. With respect to time synchronization our technique achieves timing accuracies below \SI{100}{\pico\second}, similar to other state-of-the art works presented in \cite{roehr2007method,pooler2018precise,gilligan2020white,prager2020wireless}, and with respect to frequency syntonization performance, we achieve results within a similar order of magnitude to current similar state-of-the art works presented in \cite{kenney2024frequency-synchronization, ghasemi2024time-frequency-synchronization, roehr2007method, aguilar2025joint-signal-pr} of single parts per billion.  Additional details and discussion of these experiments can be found in~\cite{merlo2025dissertation}.



\section{Conclusion}
\label{sec:conclusion}
In this work we present a novel, fully-digital high-accuracy \ac{twtt}-based joint time and frequency coordination system for \acp{cda} using unmodified \ac{cots} hardware and without relying on external time or frequency references such as \ac{gnss}; we demonstrate that the fully-digital approach shows no significant difference in performance in a system with moderate relative internode velocity; and we demonstrate an internode beamforming frequency \ac{rmse} as low as 3.73\,ppb in a dynamic case with time and phase deviations as low as \SI{60}{\pico\second} and demonstrate median coherent gains above 99\,\% with optimized coordination parameters.  This demonstrates a significant step towards fully wireless mobile \acp{cda} capable of coherent transmit and receive beamforming for the next generation of mobile wireless communication and sensing systems.

\bibliographystyle{jabbrv_ieeetr}
\bibliography{fully_digital_coordination}


\begin{IEEEbiography}[{\includegraphics[width=1in,height=1.25in,clip,keepaspectratio]{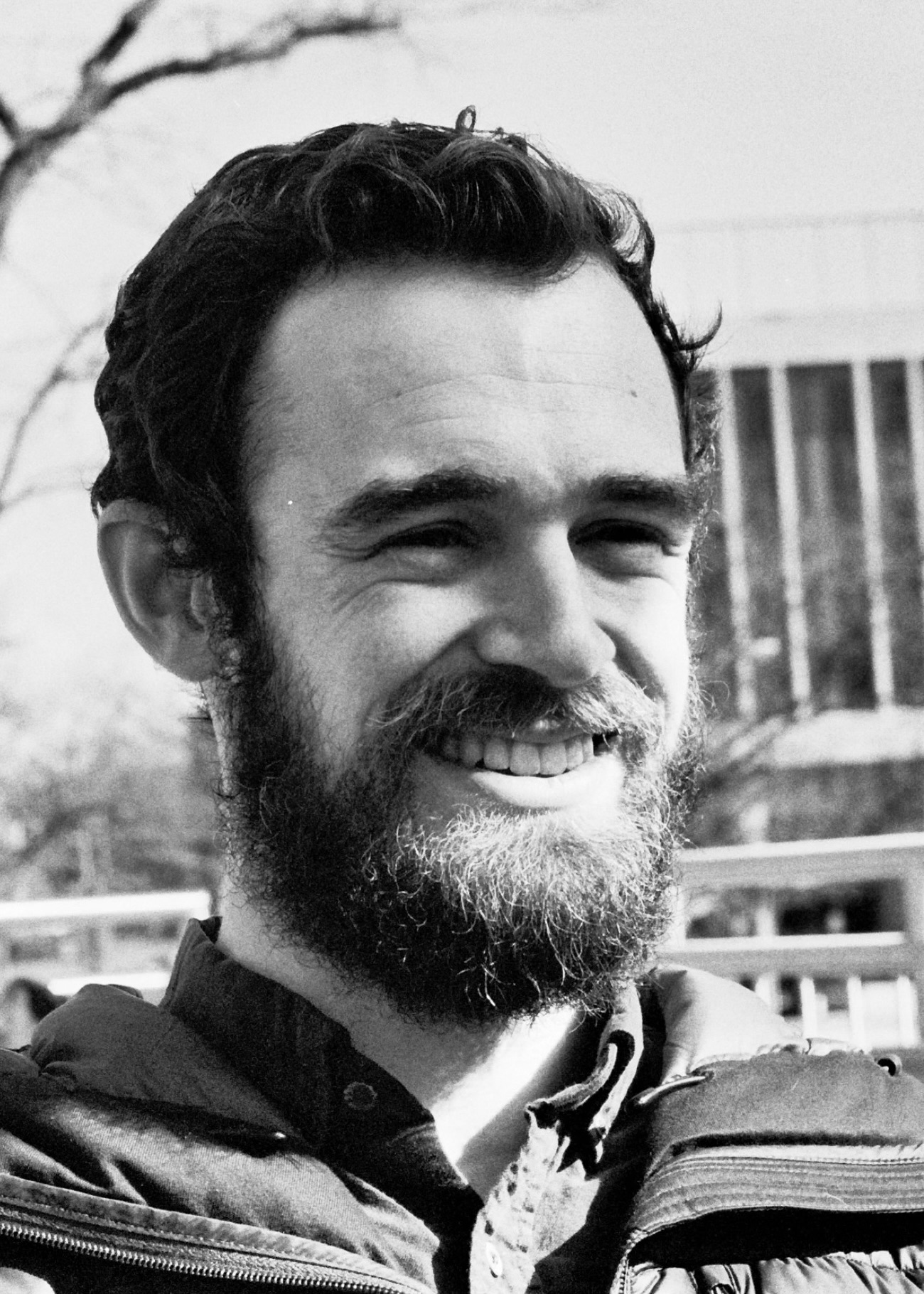}}]{Jason M. Merlo}{\space}(Member, IEEE) received the B.S. degree in computer engineering and the Ph.D. degree in electrical engineering from Michigan State University, East Lansing, MI, USA, in 2018 and 2025, respectively.
 
From 2017 to 2021, he was the project manager and the electrical systems team lead of the AutoDrive Challenge team at Michigan State University. In 2025 he joined the Department of Electrical and Computer Engineering at Michigan State University as a Fixed-Term Assistant Professor. His current research interests include software-defined radios, distributed radar and wireless systems, interferometric arrays, synthetic aperture radar, and automotive radar.
 
Dr. Merlo was a recipient of the 2023 IEEE MTT-S Graduate Fellowship and the 2023 URSI Young Scientist Award. In 2024 he placed second in the Student Paper Competition at the IEEE International Symposium on Phased Array Systems \& Technology. He was a Finalist in the Student Paper Competition at the IEEE/MTT-S International Microwave Symposium in 2024 and won First Place in 2023. He was also a Finalist and Honorable Mention in the AP-S Student Paper Competition at the IEEE International Symposium on Antennas and Propagation in 2023 and 2024, respectively. He won Second Place in the AP-S Student Design Competition at the 2020 IEEE International Symposium on Antennas and Propagation and was a finalist in the Student Safety Technology Design Competition at the 2017 International Technical Conference on Enhanced Safety of Vehicles.
\end{IEEEbiography}
\vskip 0pt plus -1fil

\begin{IEEEbiography}[{\includegraphics[width=1in,height=1.25in,clip,keepaspectratio]{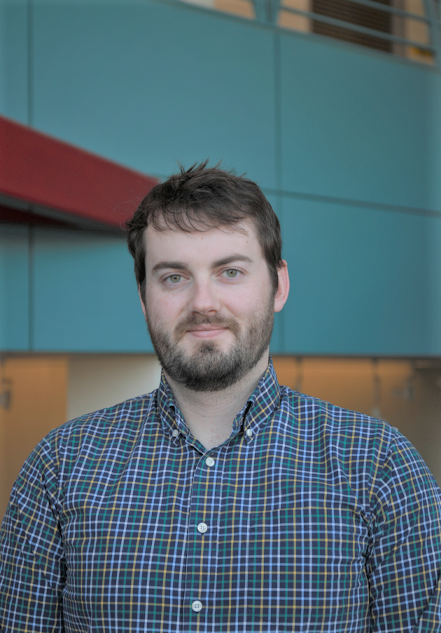}}]{Samuel Wagner} received the B.S. and Ph.D. degrees in electrical engineering from University of California, Davis, Davis, CA, USA in 2017 and 2022, respectively. 

He is currently working at Lawrence Livermore National Laboratory in Livermore, CA, USA with research focusing on high power microwaves, pulsed ultrawideband radar systems, and RF system modeling.
\end{IEEEbiography}
\vskip 0pt plus -1fil

\begin{IEEEbiographynophoto}{John B. Lancaster} received the B.S. degrees in mathematics and physics from University of Kansas, Lawrence, KS, USA in 2006 and the M.S. and Ph.D. degrees in physics from University of Missouri -- Kansas City, Kansas City, MO, USA, in 2017 and 2023, respectively.

From 2008 to 2014 he was with Horizon Analog as a principal engineer researching non-linear signal processing for positioning and communication protocols. During 2019 was the assistant director of research for the Missouri Institute of Defense and Energy developing electromagnetic compatibility approaches. He is currently a member of technical staff at Lawrence Livermore National Laboratory where his research interests are in the area of distributed phased arrays, non-linear radar, and electromagnetic compatibility.
\end{IEEEbiographynophoto}
\vskip 0pt plus -1fil

\begin{IEEEbiography}[{\includegraphics[width=1in,height=1.25in,clip,keepaspectratio]{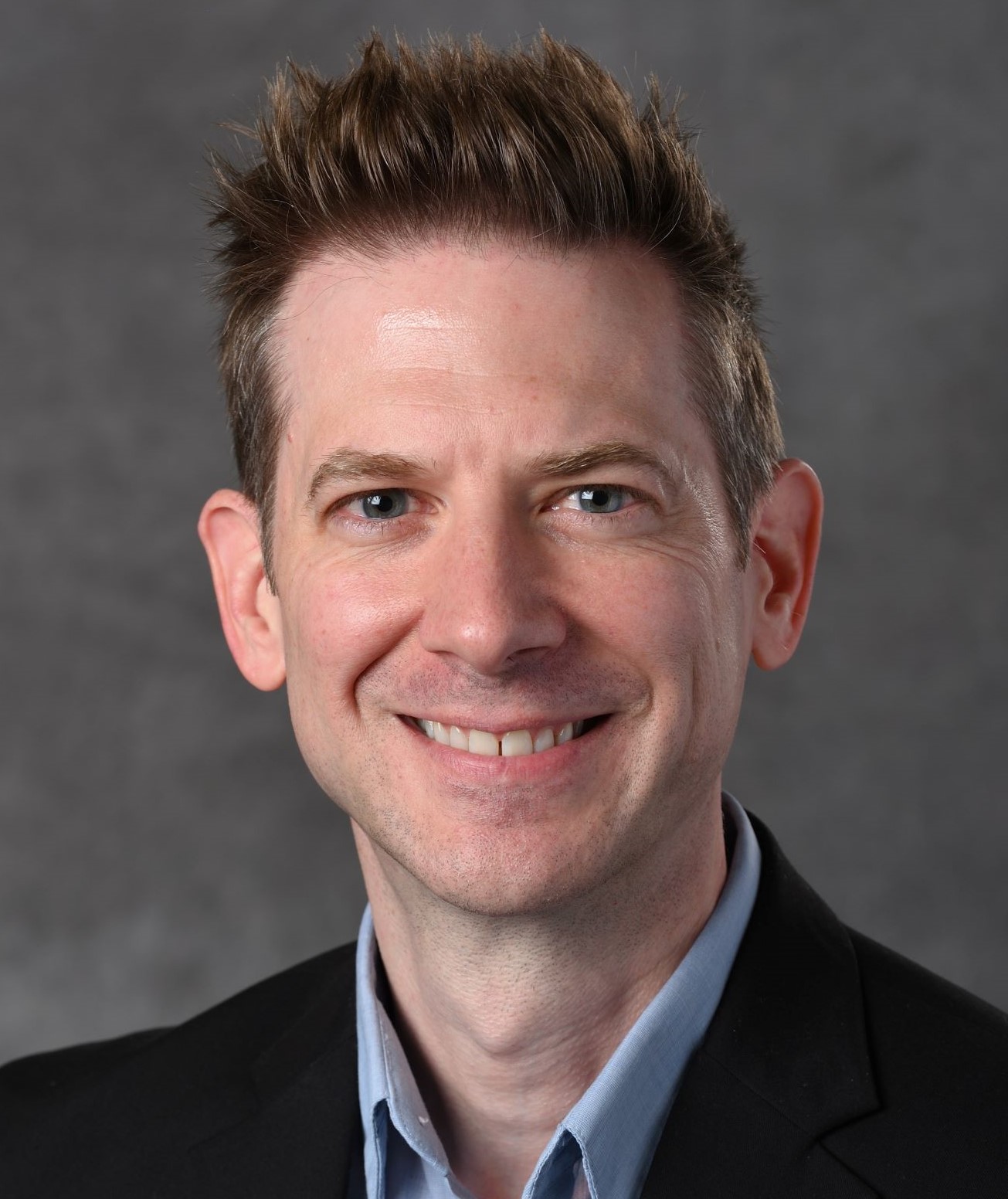}}]{Jeffrey A. Nanzer} (Senior Member, IEEE) received the B.S. degrees in electrical engineering and in computer engineering from Michigan State University, East Lansing, MI, USA, in 2003, and the M.S. and Ph.D. degrees in electrical engineering from The University of Texas at Austin, Austin, TX, USA, in 2005 and 2008, respectively.

From 2008 to 2009 he was with the University of Texas Applied Research Laboratories in Austin, Texas as a Post-Doctoral Fellow designing electrically small HF antennas and communications systems. From 2009 to 2016 he was with the Johns Hopkins University Applied Physics Laboratory where he created and led the Advanced Microwave and Millimeter-Wave Technology Section. In 2016 he joined the Department of Electrical and Computer Engineering at Michigan State University where he held the Dennis P. Nyquist Assistant Professorship from 2016 through 2021. He is currently a Professor. He directs the Electromagnetics Laboratory, which consists of the Antenna Laboratory, the Radar Laboratory, and the Wireless Laboratory. He has published more than 250 refereed journal and conference papers, two book chapters, and the book Microwave and Millimeter-Wave Remote Sensing for Security Applications (Artech House, 2012). His research interests are in the areas of distributed phased arrays, dynamic antenna arrays, millimeter-wave imaging, remote sensing, millimeter-wave photonics, and electromagnetics.

Dr. Nanzer is a member of the IEEE Antennas and Propagation Society Education Committee and the USNC/URSI Commission B. 
He was a founding member and the First Treasurer of the IEEE APS/MTT-S Central Texas Chapter, served as the Vice Chair for the IEEE Antenna Standards Committee from 2013 to 2015, and served as the Chair of the Microwave Systems Technical Committee (MTT-16), IEEE Microwave Theory and Techniques
Society from 2016 to 2018.
He was a Distinguished Microwave Lecturer for the IEEE Microwave Theory and Techniques Society (Tatsuo Itoh Class of 2022-2024) and a Guest Editor of the Special Issue on Radar and Microwave Sensor Systems in the IEEE Microwave and Wireless Components Letters in 2022. 
He was a recipient of the Withrow Junior Distinguished Scholar Award in 2024, the Google Research Scholar Award in 2022 and 2023, the IEEE MTT-S Outstanding
Young Engineer Award in 2019, the DARPA Director’s Fellowship in 2019, the National Science Foundation (NSF) CAREER Award in 2018, the DARPA Young
Faculty Award in 2017, and the JHU/APL Outstanding Professional Book Award in 2012.  
\end{IEEEbiography}

\makeatletter\@input{aux.tex}\makeatother

\end{document}